%% file: main.tex
\documentclass[5p,times,twocolumn,12pt]{elsarticle}

\usepackage{amssymb}
\usepackage{amsmath}

\usepackage{svg}
\usepackage{tikz}
\usepackage{pgfplots}
\usepackage{environ}
\usepackage[ruled]{algorithm2e} \SetKwComment{Comment}{/* }{ */}

\usepackage{acro}
\usepackage{accents}
\usepackage{subfig}
\usepackage{amsthm}

\newtheorem{thm}{Theorem}

\newtheorem{prop}[thm]{Proposition}

\theoremstyle{definition}
\newtheorem{defn}{Definition}[section]

\newtheorem{prob}{Problem}
\newtheorem{assum}{Assumption}

\usepackage{etoolbox}
\input{acronyms}

\pgfmathdeclarefunction{gauss}{2}{%
  \pgfmathparse{1/(#2*sqrt(2*pi))*exp(-((x-#1)^2)/(2*#2^2))}%
}
\pgfmathsetmacro{\gaussianY}{1/(2*sqrt(2*pi))*exp(-((-3-0)^2)/(2*2^2))}

\journal{Robotics and Autonomous Systems}
\begin{document}
\begin{frontmatter}

\title{Multi-Hypotheses Ego-Tracking for Resilient
Navigation} 
\author[DTU]{Peter Iwer Hoedt Karstensen}
\author[DTU]{Roberto Galeazzi}
\affiliation[DTU]{organization={Technical University of Denmark},
            city={Kgs. Lyngby},
            postcode={2800},
            country={Denmark}}

\begin{abstract}
Autonomous robots relying on \ac{rf}-based localization such as \ac{gnss}, \ac{uwb}, and 5G \ac{isac} are vulnerable to spoofing and sensor manipulation. This paper presents a resilient navigation architecture that combines multi-hypothesis estimation with a Poisson binomial windowed-count detector for anomaly identification and isolation. A state machine coordinates transitions between operation, diagnosis, and mitigation, enabling adaptive response to adversarial conditions. When attacks are detected, trajectory re-planning based on differential flatness allows information-gathering maneuvers minimizing performance loss. Case studies demonstrate effective detection of biased sensors, maintenance of state estimation, and recovery of nominal operation under persistent spoofing attacks.
\end{abstract}




\end{frontmatter}

\section{Introduction}

The study of resilient autonomous robotic vehicles, such as unmanned aerial vehicles, against cyber-attacks and spoofing has gained significant attention in recent years. These vehicles can be deployed for security surveillance \cite{Scherer2020, Scherer2020a, Guerra2020}, which involves coverage and patrolling tasks. Resilient navigation is a core requirement for such systems to operate in a secure and reliable manner and remains a key challenge before widespread deployment.

Redundant sensors and measurements are a prerequisite for making a system resilient against spoofing and cyber-attacks, as they enable cross-validation of measurements using a priori knowledge of the vehicle’s pose. The vehicle is typically equipped with an \ac{rf} communication device to facilitate information sharing with peers in a network and with an operator. \ac{uwb} and 5G—and beyond \ac{isac}—technologies provide accurate localization measurements in addition to communication. With multi-antenna arrays at both the transmitter and receiver sides, the receiver can determine its position relative to a single tag, base station, or peer within the network \cite{Buehrer2018, Ge2023, gonzalez-prelcic_integrated_2024}.

This paper addresses the challenges related to detecting cyber-attacks and spoofing targeting a robotic vehicle with redundant measurements used for pose estimation. Any of the measurements provided through \ac{rf} technologies such as \ac{gnss}, \ac{uwb}, or 5G \ac{isac} can be subject to attack. The robot is equipped with an exteroceptive sensor, such as a camera, to make observations of landmarks that provide additional positional information when required. The proposed approach formulates an architecture and algorithm capable of handling any form of attack that modifies measurements with the intent of corrupting the vehicle’s pose estimate. The focus is on the detection of coordinated attacks \cite{niazi_resilient_2023}.

The proposed architecture and algorithm divide measurement sources into subsets, thereby generating multiple hypothesized vehicle states. These hypotheses, together with the navigational capability of the platform, are exploited to actively seek new information to resolve potential cyber-attacks. The overall system architecture is illustrated in Fig.~\ref{fig:system_block_diagram}.

\subsection{Contributions}

The main contributions of this paper are as follows:

\begin{enumerate}
    \item An algorithm that detects abnormal sensor measurements and dynamically groups them based on the hypotheses they collectively support, ensuring resilient performance even in the presence of colluding cyber-attacks. The detector is based on a binary random variable determined from measurement inliers and outliers, from which a windowed count is generated to handle naturally occurring outliers.
    \item A multi-objective path re-planning formulation that mitigates the effects of malicious or abnormal behavior. The re-planning incorporates a notion of performance loss, enabling the selection of the path with minimal degradation.
\end{enumerate}

\subsection{Outline}

The remainder of this paper is organized as follows. 
Section~\ref{sec:related-work} presents related work on the detection and mitigation of attacks on \ac{gnss}, \ac{uwb}, and 5G \ac{isac} sensors. 
The problems addressed in this paper and the necessary definitions are provided in Section~\ref{sec:problem-statement}. 
Section~\ref{sec:sys-desc} describes the considered dynamic and measurement models. 
The system architecture and state machine enabling hypothesis generation are outlined in Section~\ref{sec:multi-hypothesis-ego-tracking}. 
Section~\ref{sec:detection} details the detection mechanism based on the windowed count. Section~\ref{sec:traj-controller} introduces the controller to track the nominal trajectory, while Section~\ref{sec:malicious-anomaly-mitigation} explains the corresponding mitigation strategy. 
Section~\ref{sec:case-studies} demonstrates the sensitivity of the detection algorithm and showcases the complete detection and mitigation approach proposed in this work. 
A case study is provided in Section~\ref{sec:case-studies} followed by a discussion on the underlying assumptions and the efficacy of the method. 
Finally, the conclusions are drawn in Section~\ref{sec:conclusion}.

\begin{figure*}[t]
    \centering
    \includegraphics[width=\linewidth]{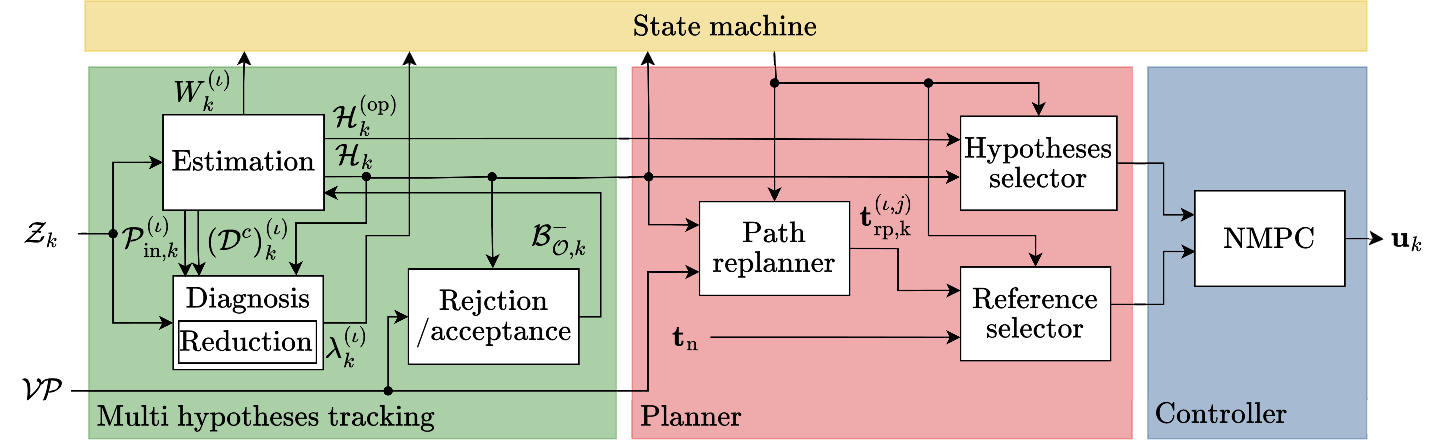}
    \caption{Block diagram of the proposed cyber-attack-resilient system architecture.}
    \label{fig:system_block_diagram}
\end{figure*}

\section{Related Work}
\label{sec:related-work}

One of the primary challenges in deploying autonomous robotic platforms lies in ensuring system security against faults and cyberattacks. A major research focus has been the detection of \ac{gnss} spoofing, as civilian \ac{gnss} services lack encryption. Efforts have concentrated on developing algorithms that enhance resilience \cite{dagdilelis_cyber-resilience_2022, yoon_towards_2019, sun_new_2021} or on extending the sensor suite with additional \ac{rf} transceivers to provide independent information channels \cite{kassas_i_2022}. In \ac{rf}-based sensing, cyberattacks may compromise either the ranging information or the inferred position of the receiver \cite{kassas_i_2022, singh_uwb-ed_2019}. Existing work often assumes that a subset of sensors or network variables is secure, frequently relying on the integrity of the cellular network \cite{kassas_i_2022}, while little attention has been given to colluding attacks across sensing domains.

\ac{rf} technologies for sensing, such as \ac{uwb} \cite{Ridolfi2021} and 5G \ac{isac} \cite{ge_integrated_2023, wymeersch_radio_2022}, have become increasingly prevalent. Although recent standards address security aspects \cite{mandyam_uwb_2022}, vulnerabilities such as the STS and Ghost attacks continue to be identified \cite{li_secure_2022, leu_ghost_2022}. 

Several studies have investigated spoofing detection and mitigation for specific \ac{rf} sensing scenarios. Guerrero-Higueras \textit{et al.}~\cite{guerrero-higueras_detection_2018} evaluated machine learning methods from Scikit-learn for a four-tag \ac{uwb} setup, though the attack types were not clearly specified. Salimpour \textit{et al.}~\cite{salimpour_exploiting_2023} proposed an iterative scheme for fully connected networks to identify and exclude malicious nodes. Outlier rejection based on the \ac{md} remains a common approach for improving robustness \cite{dwek_improving_2020, hol_tightly_2009}. Chen \textit{et al.}~\cite{chen_spoofing_2024} addressed compromised anchors using an active–passive ranging strategy, allowing other anchors to infer their distance to the affected node. Venturino \textit{et al.}~\cite{venturino_adaptive_2024} developed a change detection method using \ac{aoa} signals from a \ac{gnss} antenna array and an \ac{imu}, although the method is sensitive to other sensors skewing the vehicle pose. He \textit{et al.}~\cite{he_resilient_2023} modeled a spoofed \ac{gnss} receiver and forged control inputs using a game-theoretic framework, assuming noiseless sensors.

The field of \ac{fd} is mature, with classical model-based methods employing analytical redundancy relations \cite{staroswiecki_analytical_2001} and banks of observers \cite{willsky_survey_1976, frank_fault_1990}. These rely on statistical change detection and enable fault isolation under known fault profiles. However, in adversarial environments, faults stem from spoofing or cyber attacks, where the attacked sensors are unknown.

Multiple-hypothesis navigation has long been used in simultaneous localization and mapping \cite{Arras2002}. Robots test multiple pose hypotheses based on non-unique map features. Colle \textit{et al.}~\cite{colle_multihypothesis_2017} employed a set-based approach to manage sporadic sensor outliers, while Niazi \textit{et al.}~\cite{niazi_resilient_2023} extended this concept to detect cyberattacks on arbitrary subsets of sensors using specialized intersection and union rules, albeit with high computational complexity. Jurado \textit{et al.}~\cite{jurado_residual-based_2020} proposed a bank of \ac{kf}s excluding different sensors to isolate faults, and Gipson \textit{et al.}~\cite{gipson_resilience_2022} later refined this by revalidating sensors and discussing guarantees on state observability.

\section{Problem Statement}
\label{sec:problem-statement}

Consider a mobile robot maneuvering within a surveillance area. The environment contains a sparse set of a priori known landmarks with associated viewpoints and a set of \ac{rf} anchors. The robot follows a nominal patrol trajectory.

The robot is equipped with exteroceptive sensors that enable accurate self-localization and navigation. These sensors generate measurements associated with identifying tags of the corresponding anchors or landmarks. The viewpoints contain unique features that allow them to be labeled with distinct tags.

Some sensors are more susceptible to attacks than others, depending on the medium through which measurements are acquired. As summarized in \cite{xu_sok_2023}, the feasibility of an attack depends on the required equipment size and cost, the attacker’s ability to remain concealed, and the operational range of the attack. For \ac{rf} sensors such as \ac{gnss} receivers and communication devices capable of ranging, an attacker can inject erroneous information into the signals from a safe distance. In contrast, camera-based sensors rely on physical features, requiring an attacker to interfere within the surveillance area. For example, an adversary may place counterfeit ArUco markers to mislead the system. In this work, attacks are assumed to target only \ac{rf}-based measurement sources.
\begin{defn}
    A measurement source is considered \textit{attacked} when it provides erroneous information. A measurement is \textit{spoofed} when $\boldsymbol{\epsilon}_k^{(i)}$ in Eq.~\eqref{eq:measurement_func} is non-zero. 
\end{defn}
The signal $\boldsymbol{\epsilon}_k^{(i)}$ is inherently random from the perspective of the robot, and no assumptions are made about its statistics or the attack probability.
\begin{prob}
    A robot operates in an adversarial surveillance area where its \ac{rf} transceivers, and consequently its navigational capability, may be compromised. The objective is to distinguish truthful from falsified information using redundant measurements. 
\end{prob}
\begin{prob}
    Following an attack on measurement sources, the robot may face multiple plausible pose hypotheses corresponding to uncertain locations.
\end{prob}
The analysis in this paper is based on the following assumptions:
\begin{assum}
    \label{assum:known-natur-outlier-prob}
    Sensors may generate outliers, and the probability of these outliers is known and used by the windowed count detector. The local modes are well separated from the global mode.
\end{assum}
\begin{assum}
    \label{assum:E=True}
    The expected value of the estimated states equals the true state, used in the design of the windowed count detector.
\end{assum}
\begin{assum}
    \label{assum:persistent-attack}
    An attack, once initiated, persists over time, indicating that the attacker aims to continuously disrupt the system.
\end{assum}
\begin{assum}
    The surveillance area is convex in the configuration space, allowing the robot to maneuver freely without collision constraints.
\end{assum}

In contrast to the approaches in \cite{jurado_residual-based_2020} and \cite{gipson_resilience_2022}, which employ fixed banks of observers and rely on discarding sensors until validation is achieved, the proposed framework in this paper builds the bank of filters, referred to as hypothesis in this paper, dynamically. Instead of permanently excluding a measurement source once suspected, each hypothesis continues to track its associated subset of sensors. This design choice preserves information that later will become useful and enables a seamless transition into the mitigation phase, where maintaining multiple active hypotheses allows re-evaluation and recovery of previously disregarded measurement sources.

\subsection{Notation}
The subscript $k$ denotes the discrete time index. The symbols $k-$ and $k+$ represent quantities immediately before and after an operation at time step $k$, respectively. The superscripts $\iota$ and $\nu$ are used as indices for hypotheses, whereas the superscript $(i)$ refers to sensor measurements unless stated otherwise. The operator $\uplus$ denotes a disjoint union. Definitions of all variables used throughout this paper are provided in Tables~\ref{tab:variables} and \ref{tab:variables_cont}.

The likelihood of a Gaussian distribution with parameters $\boldsymbol{\mu} \in \mathbb{R}^{n_z}$ and covariance matrix $\mathbf{R}$, evaluated at a given point $\mathbf{z}$, is expressed as
\begin{equation}
    \begin{split}
        \ell_{N(\boldsymbol{\mu}, \mathbf{R})}(\mathbf{z}) =  
        \frac{1}{\sqrt{2\pi}^{n_z^{(i)}}\sqrt{|\mathbf{R}|}}\mathrm{e}^{-\frac{1}{2}\left(\boldsymbol{\mu}-\mathbf{z}\right)^T\left(\mathbf{R}\right)^{-1}\left(\boldsymbol{\mu}-\mathbf{z}\right)}
    \end{split}
    \label{eq:likelihood}
\end{equation}

\begin{table}[]
    \centering
    \caption{Explanation of symbols used throughout the paper}
    \label{tab:variables}
    \begin{tabular}{cp{0.7\columnwidth}}
    \hline
    Variable & Definition                                            \\ \hline
     $\mathcal{RF}$ & Set of \ac{rf} anchor locations \\
     $\mathcal{VP}$ & Set of view point locations \\
     $\mathbf{t}_{\mathrm{n}}$ & Nominal trajectory \\
     $\mathcal{Z}_k$ & Set of measurements at $k$ \\
     $s^{(i)}$ & Tag of measurement $i$ \\
     $\mathcal{O}_k$ & Collection of all tags at $k$ \\
     $\boldsymbol{\epsilon}_k^{(i)}$ & Time-dependent attack variable affecting measurement $i$\\
     $W$ & Window size for detection algorithm \\
     $\mathcal{H}_k$ & Collection of all hypotheses at $k$ \\
     $h_k^{(\iota)}$ & Hypothesis $\iota$ at $k$ \\
     $h_k^{(\mathrm{op})}$ & Operational hypothesis at $k$ \\
     $\lambda_k^{(\iota)}$ & Alarm signal of hypothesis $\iota$ \\
     $W_k^{(\iota)}$ & Existence count of hypothesis $\iota$ at $k$ \\
     $\alpha_\mathrm{d}$ & Covariance matrix scaling factor \\
     $\mathcal{P}_{\mathrm{in},}^{(\iota)}$ & Collection of all inlier probabilities over $W_k^{(\iota)}$ \\
     $ \left(\mathcal{D}^c\right)_{\mathrm{in},}^{(\iota)}$ & Collection of all outlier counts over $W_k^{(\iota)}$ \\
     $\mathcal{B}_{\mathcal{O},k}^-$ & Collection of all blacklisted measurement sources at $k$ \\
     $\mathbf{x}_k$ & True state of robot \\
     $\mathbf{u}_k$ & Robot actuator input \\
     $T_s$ & Sampling period \\
     $\mathbf{w}_{\mathrm{d},k}$ & Process noise \\
     $\mathbf{w}_{\mathrm{m},k}^{(i)}$ & Measurement noise of source $i$ \\
     $\mathbf{p}$ & Position vector \\
     $\mathbf{q}$ & Pose vector \\
     $\mathbf{P}$ & State error covariance matrix \\
     $\mathbf{R}$ & Measurement noise covariance matrix \\
     $\alpha_\chi$ & $\chi^2$ Percentile used to determine $\gamma_{\alpha_\chi}$ \\
     $\gamma_{\alpha_\chi}$ & Threshold determined based on percentile $\alpha_\chi$  \\
     $\alpha_\mathrm{F}$ & Percentile for determining like hypotheses \\
     $W_\mathrm{p}$ & Window for determining like hypotheses\\
     \hline      
    \end{tabular}
\end{table}

\begin{table}[]
    \centering
    \caption{Table \ref{tab:variables} continued.}
    \label{tab:variables_cont}
    \begin{tabular}{cp{0.7\columnwidth}}
    \hline
    Variable & Definition                                            \\ \hline
     $\tau_{\mathrm{r}/\mathrm{a}}$ & Number of time steps for residing in view point region\\
     $M_\mathrm{c}$ & NMPC prediction horizon\\
     $M_\mathrm{s}$ & Prediction step selected for path re-planning\\
     $M_\mathrm{rp}$ & Path re-planning prediction horizon\\
     $\mathbf{W}_{\mathrm{VP}}$ & Weight matrix on VP re-planning cost\\
     $\mathbf{W}_{g}$ & Weight matrix on to-go cost\\
     $\mathbf{W}_{f}$ & Weight matrix on final cost \\
     \hline          
    \end{tabular}
\end{table}

\section{Cyber-Physical Resilient System Architecture}
\label{sec:sys-desc}

A mobile robot operating in an adversarial environment may experience attacks on its sensors. Since any measurement source can be compromised, truthful information must be discerned from falsified data. This challenge resembles sensor isolation in \ac{fd}, where specific faulty sensors are typically anticipated and handled through isolability analysis and fault-tolerant control. In contrast, under adversarial conditions, no prior knowledge exists about which sensors may be attacked. 

To address this, the proposed architecture maintains multiple hypotheses, each relying on a subset of measurement sources—conceptually similar to observer banks in \ac{fd}. The robot re-plans its motion to gather additional information through exteroceptive sensors (e.g., a camera) and incrementally rejects inconsistent hypotheses until a single consistent one remains. The overall architecture, shown in Fig.~\ref{fig:system_block_diagram}, extends the classical sensor fusion–planning–control structure with a dedicated state machine, described in Section~\ref{sec:state-machine}.

\subsection{Agent \& Measurement Model}
The robot dynamics are described by the nonlinear differential equation
\begin{equation}
    \dot{\mathbf{x}}(t) = \mathbf{f}(\mathbf{x}(t), \mathbf{u}(t))
    \label{eq:system_model}
\end{equation}
where $\mathbf{x}(t)$ is the state vector, $\mathbf{u}(t)$ the control input, and $\mathbf{f}(\cdot)$ a nonlinear function representing the vehicle dynamics. Discretization with sampling period $T_s$ yields
\begin{equation}
    \mathbf{x}_{k+1} = \mathbf{f}(\mathbf{x}_k, \mathbf{u}_k) + \mathbf{w}_{\mathrm{d},k}
    \label{eq:system_model_disc}
\end{equation}
where $\mathbf{w}_{\mathrm{d},k} \sim N(\mathbf{0}, \mathbf{Q}_k)$ represents process noise. The inputs $\mathbf{u}_k$ are measured via an \ac{imu} with covariance matrix denoted as $\mathbf{Q}_\mathrm{IMU}$.

The configuration space is two-dimensional, though the formulation generalizes to three dimensions. The robot’s position and heading are $\mathbf{p}_k = [x_k, y_k]^{T}$ and $\theta_k$, respectively, with pose $\mathbf{q}_k = [\mathbf{p}_k^{T}, \theta_k]^{T}$.

Sensor measurements are subject to natural outliers characterized by varying density regions. The $N_{\mathrm{S},k}$ measurements at time step $k$ are modeled as
\begin{equation}
    \mathbf{z}_k^{(i)} = \mathbf{h}^{(i)}(\mathbf{x}_k + \mathbf{x}_{k}^{(l)}, \boldsymbol{\epsilon}_k^{(i)}, e^{(i)}) + \mathbf{w}^{(i,l)}_{\mathrm{m},k} \; , \quad l\sim\mathrm{Cat}(p_0,\dots,p_L)
    \label{eq:measurement_func}
\end{equation}
where $\mathbf{h}^{(i)} = \mathbf{h}_{S(s^{(i)})}(\cdot)$ maps the state to the measurement space based on tag $s^{(i)}$. The noise term $\mathbf{w}_{\mathrm{m},k}^{(i,l)} \sim N(\mathbf{0}, \mathbf{R}_k^{(i,l)})$ corresponds to the $l$-th Gaussian component with weight $p_l$, $\sum_{l=0}^{L} p_l = 1$. The variable $e^{(i)}$ represents sensor-specific parameters for example anchor or viewpoint positions, and $\boldsymbol{\epsilon}_k^{(i)}$ denotes a time-varying adversarial signal that corrupts the measurement.

\subsection{Measurement Sources}
\label{sec:sec:sensor-suite}

The robot is equipped with a camera, a \ac{gnss} receiver, and a communication device capable of providing range and bearing information, such as an \ac{uwb} sensor with an antenna array. The \ac{gnss} receiver provides a single global measurement, while \ac{rf} anchors with known locations $\mathbf{p}_\mathrm{RF}^{(j)} \in \mathcal{RF}$ enable local ranging. Communication is possible only when anchors are within range $\rho_{\mathrm{RF}}$, meaning that only a subset of anchors contributes measurements at each time step $k$. The camera observes landmarks with known poses $\mathbf{q}_{\mathrm{VP}}^{(j)} \in \mathcal{VP}$, such as ArUco markers or unique semantic objects, visible within specific distance and orientation constraints.

All measurements are collected in the set $\mathcal{Z}_k = \{\mathbf{z}_k^{(1)}, \dots, \mathbf{z}_k^{(N_{\mathrm{S},k})}\}$ with corresponding tags $\mathcal{O}_k = \{s^{(1)}, \dots, s^{(N_{\mathrm{S},k})}\}$. A measurement source is \textit{blacklisted} when deemed untrustworthy, and the set of such sources is denoted $\mathcal{B}_{\mathcal{O},k}^{-}$, as further described in Section~\ref{sec:malicious-anomaly-mitigation}.

\section{State Machine}
\label{sec:state-machine}

\begin{figure}[t]
    \centering
    \includegraphics[width=0.6\linewidth]{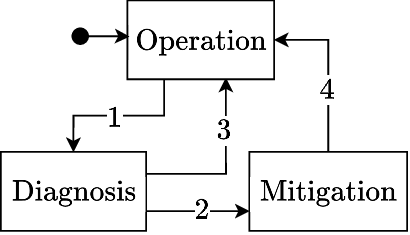}
    \caption{System state machine and transitions.}
    \label{fig:state-machine}
\end{figure}

The state machine, illustrated in Fig.~\ref{fig:state-machine}, comprises three states and four possible transitions. It determines whether the robot continues along the nominal path or initiates path re-planning to acquire additional information. The transition logic, expressed in terms of the symbols introduced in Section~\ref{sec:multi-hypothesis-ego-tracking}, is summarized below.

\subsection{Operation and diagnosis state}

Upon initialization, the robot enters the \textit{Operation} state, representing both nominal and degraded operation modes. In this state, the robot estimates a set of hypotheses supported by subsets of measurement sources. Transitions into this state from either the \textit{Diagnosis} or \textit{Mitigation} state occur when a previous alarm is determined to be false or when an attack has been successfully mitigated.

The robot transitions to the \textit{Diagnosis} state when the diagnosis module raises an alarm. In this state, the robot updates existing hypotheses and generates new ones when measurements are inconsistent across sources. The currently active hypothesis is used for nominal path tracking. The multi-hypothesis ego-tracking process is detailed in Section~\ref{sec:multi-hypothesis-ego-tracking}.

\subsection{Mitigation state}
\label{sec:mitigation}

In the \textit{Mitigation} state, measurement sources have been partitioned into disjoint subsets, each defining a hypothesis. The robot re-plans its trajectory to gather additional evidence that enables acceptance or rejection of these hypotheses. Re-planning is guided by the goal of visiting informative viewpoints: measurements that confirm a single hypothesis are retained, while absent or inconsistent measurements lead to hypothesis rejection. The path re-planning process is described in Section~\ref{sec:malicious-anomaly-mitigation}. During this state, only the removal of measurement sources is permitted, accounting for those that move out of range.

\section{Multi hypothesis ego tracking}
\label{sec:multi-hypothesis-ego-tracking}

This section describes the estimation, diagnosis, and hypothesis management modules shown in Fig.~\ref{fig:est_diag_flow}. All symbols and variables introduced here are later used to define the state machine transitions.

\subsection{Hypotheses}
\label{sec:sec:hypo}

At time step $k$, the set of hypotheses is
\begin{equation}
    \mathcal{H}_{k} = \left\{h^{(\iota)}_k\right\}_{\iota=1}^{|\mathcal{H}_{k}|} \quad h^{(\iota)}_k = \left(\left(\boldsymbol{\mu}_{(\cdot)}^{(\iota)},\mathbf{P}_{(\cdot)}^{(\iota)}\right),\mathcal{O}_k^{(\iota)}\right)
\end{equation}
where $\boldsymbol{\mu}_{(\cdot)}^{(\iota)}$ and $\mathbf{P}_{(\cdot)}^{(\iota)}$ correspond to the parameters defined in Eqs.~\eqref{eq:predicted}–\eqref{eq:predicted_non_infla}, and $\mathcal{O}_k^{(\iota)}\subseteq\mathcal{O}_k$ denotes the measurement sources supporting $h^{(\iota)}_k$. Each hypothesis has an associated alarm $\lambda_k^{(\iota)}\in\{0,1\}$, a collection of unlikely measurement probabilities $\mathcal{P}_{\mathrm{in},k}^{(\iota)}$, counts $\left(\mathcal{D}^c\right)^{(\iota)}_{k}$, and an existence counter $W_k^{(\iota)}$. The generation of these quantities and alarm logic are described in Section~\ref{sec:detection}. 

\begin{figure}[!t]
    \centering
    \includegraphics[width=0.7\linewidth]{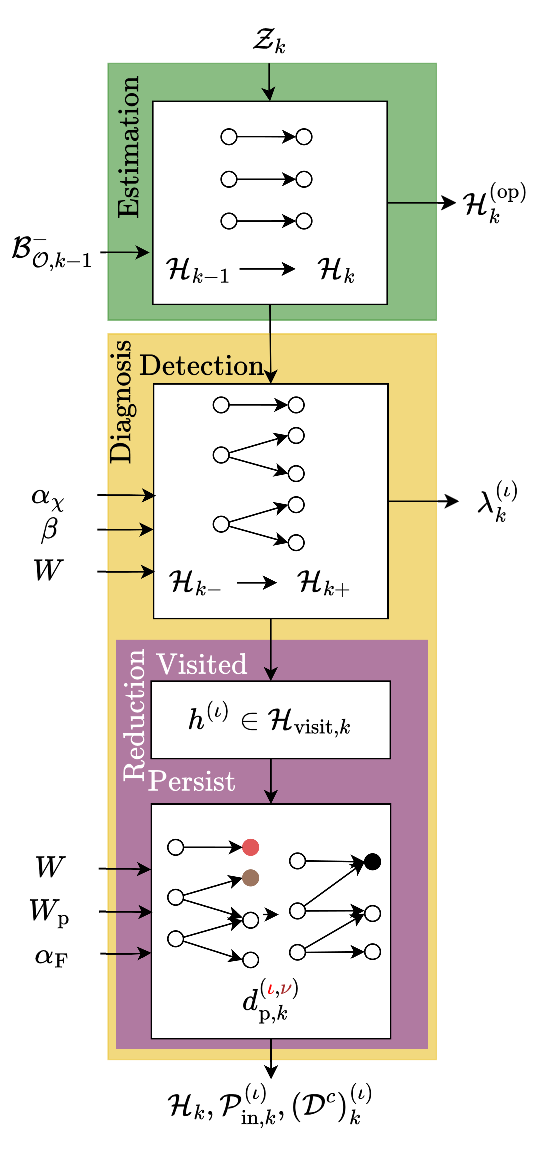}
    \caption{Diagram depicting the operations in a chronological order and how the parameters enter into the algorithm. The vertical arrows contain $\mathcal{H}_k$, $\mathcal{P}_{\mathrm{in},k}^{(\iota)}$, $\left(\mathcal{D}^c\right)^{(\iota)}_{k}$ and $W_k^{(\iota)}$.}
    \label{fig:est_diag_flow}
\end{figure}

When the detection algorithm raises $\lambda_{k_d}^{(\iota)}=\{1\}$ for one or more hypotheses, new hypotheses are generated as
\begin{equation}
    \begin{split}
        \mathcal{H}_{k_d+}^{(1)} &= \left\{\left(\left(\boldsymbol{\mu}_{(\cdot)}^{(\iota)},\;\alpha_\mathrm{d}\mathbf{P}_{(\cdot)}^{(\iota)}\right),\;\mathcal{O}_{k_d+}^{(\nu)}\right)\;\Big|\; \right.\\
        &\qquad\left. \vphantom{\Big|} \mathcal{O}_{k_d+}^{(\nu)}\subset\mathcal{O}_{k_d-}^{(\iota)},\; |\mathcal{O}_{k_d+}^{(\nu)}|=|\mathcal{O}_{k_d-}^{(\iota)}|-1\right.\\
        &\qquad\left. \vphantom{\Big|} \mathcal{O}_{k_d+}^{(\nu)} \notin \mathcal{O}_{k_d-}, \mathcal{O}_{k_d+}^{(\nu)} \not\subset \mathcal{O}_{k_d+}^{(\iota)},\right.\\
        &\qquad\left.\vphantom{\Big|} \lambda_{k_d-}^{(\iota)}=\{1\},\; \iota=0,\dots|\mathcal{H}_{k_d-}| \right\}
    \end{split}
    \label{eq:new_hypo}
\end{equation}
where $\alpha_\mathrm{d}>1$ inflates the covariance of inherited densities. The associated probabilities, counts, and counters are reset. Hypotheses with $\lambda_{k_d-}^{(\iota)}=\{0\}$ are denoted $\mathcal{H}_{k_d+}^{(0)}$, and the complete updated set is
\begin{equation}
    \mathcal{H}_{k_d+} = \mathcal{H}_{k_d+}^{(1)} \cup\mathcal{H}_{k_d+}^{(0)}
\end{equation}
A separate operational hypothesis $h_k^{(\mathrm{op})}$ includes all measurement sources $s^{(i,j)}\notin\mathcal{B}_{\mathcal{O},k}^-$. The overall estimation and reduction flow is depicted in Fig.~\ref{fig:est_diag_flow}.

\subsubsection{Hypotheses estimation}
State estimation employs nonlinear \ac{kf} variants such as the \ac{ekf}. The prior, predicted, and posterior densities are
\begin{subequations}
    \begin{align}
        \mathbf{\hat{x}}^{(\iota)}_{k-1|k-1} &\sim N\left(\boldsymbol{\mu}_{k-1|k-1}^{(\iota)}, \mathbf{P}_{k-1|k-1}^{(\iota)}\right) \label{eq:prior}\\
        \mathbf{\hat{x}}^{(\iota)}_{k|k-1} &\sim N\left(\boldsymbol{\mu}_{k|k-1}^{(\iota)}, \mathbf{P}_{k|k-1}^{(\iota)}\right) \label{eq:predicted}\\
        \mathbf{\hat{x}}_{k|k}^{(\iota)} &\sim N\left(\boldsymbol{\mu}_{k|k}^{(\iota)}, \mathbf{P}_{k|k}^{(\iota)}\right) \label{eq:posterior} \\
        \mathbf{\hat{z}}_{k|k-1}^{{(\iota,i)}} &\sim N\left(\boldsymbol{\mu}_{\mathbf{z},k}^{(\iota,i)},\mathbf{P}_{k|k-1}^{(\iota,i)}\right)\label{eq:predicted_non_infla}
    \end{align}
\end{subequations}
where $\boldsymbol{\mu}_{\mathbf{z},k}^{(\iota,i)}=\mathbf{h}^{(i)}(\boldsymbol{\mu}_{k|k-1}^{(\iota)},e^{(i)})$ and $\mathbf{P}_{k|k-1}^{(\iota,i)}=\mathbf{H}_{k}^{(\iota,i)}\mathbf{F}_{k}^{(\iota)}\mathbf{P}_{k-1|k-1}(\mathbf{F}_{k}^{(\iota)})^{T}(\mathbf{H}_{k}^{(\iota,i)})^{T}$. The matrices $\mathbf{F}$ and $\mathbf{H}_k^{(\iota,i)}$ are Jacobians of the system \eqref{eq:system_model_disc} and measurement models \eqref{eq:measurement_func}, respectively.

A validation region is defined as
\begin{equation}
    \mathcal{E}_{\mathbf{R}}^{\gamma_{\alpha_\chi}}(\mathbf{y}) = 
    \left\{\mathbf{x} \; \middle| \; (\mathbf{y}-\mathbf{x})^T\left(\mathbf{R}\right)^{-1}(\mathbf{y}-\mathbf{x})\le\gamma_{\alpha_\chi}\right\}
    \label{eq:likely_meas_region}
\end{equation}
where $\gamma_{\alpha_\chi}$ is obtained from the inverse $\chi^2$ \ac{cdf} at percentile $\alpha_\chi$. When $\mathbf{z}_k^{(i)} \notin \mathcal{E}_{\mathbf{S}_k^{(\iota,i)}}^{\gamma_{\alpha_\chi}}\left(\boldsymbol{\mu}_{\mathbf{z},k}^{(\iota,i)}\right)$, where $\mathbf{S}_k^{(\iota,i)} = \mathbf{H}_k^{(\iota,i)}\mathbf{P}_{k|k-1}^{(\iota)}\left(\mathbf{H}_{k}^{(\iota,i)}\right)^\mathrm{T} + \mathbf{R}_k^{(i)}$, the measurement $\mathbf{z}_k^{(i)}$ is not included in the update step of $h^{(\iota)}$.

\subsubsection{Hypotheses Reduction}
During operation, previously rejected hypotheses may be revisited. The set $\mathcal{O}_{\mathrm{visit},k}$ stores the subset of tags that have been rejected. At each instance of $\mathcal{H}_{k_d+}^{(1)}$, members will be discarded when their $\mathcal{O}_k^{(\iota)}\in\mathcal{O}_{\mathrm{visit},k}$. Two hypotheses are merged when their \ac{md}
\begin{equation}
    d_{\mathrm{p},k}^{(\iota,\nu)} = \begin{cases}
        1 & \text{if } \mathrm{dist}\left(h^{(\iota)},h^{(\nu)}\right) \leq \
        F_{\chi^2}^{-1}(n_x,\alpha_\mathrm{F})\\
        0 & \text{otherwise} 
    \end{cases}
\end{equation}
satisfies a proximity criterion. Linear pooling \cite{Koliander2022} occurs if for hypotheses $\iota$ and $\nu$ when $\mathcal{O}_k^{(\iota)} \subset \mathcal{O}_k^{(i)}$ or vice versa and $\sum_{k=\kappa-W}^W d_{\mathrm{p},\kappa}^{(\iota,\nu)} \ge W_\mathrm{p}$, where $W_\mathrm{p}$ is a user-defined threshold. It also occurs when $\sum_{k=\kappa-W}^W d_{\mathrm{p},\kappa}^{(\iota,\nu)} \ge W_\mathrm{p}$ and both hypotheses $\iota$ and $\nu$ have existed for $W$ timesteps.

\subsubsection{Adding and Removing Measurement Sources}
\label{sec:sec:sec:add-remove-meas}

Due to range limitations, measurement sources may appear or disappear during operation. When a source $s^{(i)}\notin\mathcal{O}_k$, the sets are updated as $\mathcal{O}_{k}^{(\iota)} = \mathcal{O}_{k-1}^{(\iota)}\setminus s^{(i)}$ for all $\iota$. A counter function $W_{\mathcal{O},k}(s)$ tracks the duration of each source’s inclusion:
\begin{equation}
    W_{\mathcal{O},k+1}(s) = W_ {\mathcal{O},k}(s) + 1 \quad \forall s\in\mathcal{O}_k
\end{equation}

\subsubsection{State Machine Transition Logic}
\label{sec:sec:sec:merging-hypotheses-before-state-transition}
Transitions are expressed in terms of the introduced variables.  
From operation to diagnosis:
\begin{equation}
    \texttt{trans}_1 :\exists \iota \text{ where } \lambda_k^{(\iota)}=\{1\}
\end{equation}
From diagnosis to mitigation:
\begin{equation}
    \begin{split}
        \texttt{trans}_2 &: \left(\uplus_\iota \mathcal{O}_k^{(\iota)} = \mathcal{O}_k \right)\wedge\left( W_k^{(\iota)} \geq W \;\forall\iota\right) \\
        &\quad\wedge \left(W_{\mathcal{O},k}(s)\geq \frac{W}{2} \;\forall s\in\mathcal{O}_k\right) \\
    \end{split}
\end{equation}
and from diagnosis to operation:
\begin{equation}
    \begin{split}
        \texttt{trans}_3 &: \left(\uplus_\iota \mathcal{O}_k^{(\iota)} \neq \mathcal{O}_k\right) \wedge \left(W_k^{(\iota)} \geq W \;\forall \iota\right) \\
        &\quad\wedge\vphantom{W^{(\iota)}} \left(W_{\mathcal{O},k}(s)\geq \frac{W}{2} \;\forall s\in\mathcal{O}_k\right)
    \end{split}
    \label{eq:sm-diagnosis-logic}
\end{equation}

The final transition from mitigation to operation occurs when only one valid hypothesis remains:
\begin{equation}
    \texttt{trans}_4 : \left|\left\{\iota \mid \mathcal{O}_k^{(\iota)}\not\subset \mathcal{B}_{\mathcal{O},k}^-\right\}\right| = 1
\end{equation}

\subsection{View point region}
\label{seq:seq:view-point-region}
Each viewpoint is associated with an axis-aligned rectangular region
\begin{equation}
    \mathcal{F}_\mathcal{VP}\left(\mathbf{q}_\mathrm{VP}^{(i)}\right) = \left\{\mathbf{x}\mid \mathbf{q}_\mathrm{VP}^{(i)} - \mathbf{b} \leq\mathbf{x} \leq \mathbf{q}_\mathrm{VP}^{(i)} + \mathbf{b} \right\} 
    \label{eq:view-point-vicinity}
\end{equation}
where $\mathbf{b}$ defines the region size. These small regions are typically disjoint from the nominal trajectory, requiring minor detours for camera-based measurements. A binary variable $\delta_{\mathrm{ps},k}^{(\iota,i)}$ is set when $95\%$ of $C$ samples drawn from $N\left(\mathbf{q}_{k|k}^{(\iota)},\mathbf{P}_{k|k,\mathbf{q}}^{(\iota)}\right)$ lie within the region, with $\mathbf{P}_{k|k,\mathbf{q}}^{(\iota)}$ denoting the pose covariance.



\section{Detection through Outlier Counting}
\label{sec:detection}

This section describes the detection mechanism responsible for generating the alarms $\lambda_k^{(\iota)}$. The method counts occurrences where the predicted measurement $\hat{\mathbf{z}}_{k|k-1}^{(\iota,i)}$ lies outside the validation region $\mathcal{E}_{\mathbf{R}^{(i)}}^{\gamma_{\alpha_\chi}}(\mathbf{z}_k^{(i)})$ defined in Eq.~\eqref{eq:likely_meas_region}. Each event is modeled as a Bernoulli random variable:
\begin{equation}
    \begin{split}
        \delta_k^{(\iota,i)} &= \mathbf{1}_{\mathcal{E}_{\mathbf{R}^{(i)}}^{\gamma_{\alpha_\chi}}\left(\mathbf{z}_{k}^{(i)}\right)}\left(\hat{\mathbf{z}}^{(\iota,i)}_{k|k-1}\right)\\ 
        &= \begin{cases}
            1 & \text{if  } \hat{\mathbf{z}}^{(\iota,i)}_{k|k-1} \in \mathcal{E}_{\mathbf{R}^{(i)}}^{\gamma_{\alpha_\chi}}\left(\mathbf{z}_{k}^{(i)}\right) \\
            0 & \text{otherwise} 
        \end{cases}
    \end{split}
    \label{eq:bernoulli-random-var}
\end{equation}
The outlier probability for a given hypothesis is
\begin{equation}
    \begin{split}
        P_{\mathrm{out},k}^{(i,\iota)} =&\mathbb{P}\left(\hat{\mathbf{z}}_{k|k-1}^{(\iota,i)}\not\in\mathcal{E}_{\mathbf{R}^{(j)}}^{\gamma_{\alpha_\chi}}\left(\mathbf{z}_k^{(i)}\right)\mid l\right) \\
        =& \sum_{m=0}^L p_m \mathbb{P}\left(\hat{\mathbf{z}}_{k|k-1}^{(\iota,i)}\not\in\mathcal{E}_{\mathbf{R}^{(j)}}^{\gamma_{\alpha_\chi}}\left(\mathbf{z}_k^{(i)}\right)\mid l=m\right)
    \end{split}
    \label{eq:outlier-prob-general}
\end{equation}

Applying Assumption~\ref{assum:E=True} implies that $\mathbb{E}(\hat{\mathbf{z}}^{(\iota,i)}_{k|k-1}) = z_{k,\mathrm{MAP}}^{(i)}$, such that the \ac{ekf} tracks the high-density measurement region. Following Assumption \ref{assum:known-natur-outlier-prob} the measurement bias for $l>0$ is large, the above simplifies to
\begin{equation}
    \begin{split}
        P_{\mathrm{out},k}^{(i,\iota)} =&\mathbb{P}\left(\hat{\mathbf{z}}_{k|k-1}^{(\iota,i)}\not\in\mathcal{E}_{\mathbf{R}^{(j)}}^{\gamma_{\alpha_\chi}}\left(\mathbf{z}_k^{(i)}\right)\mid l\right) \\
        =& p_0 \mathbb{P}\left(\hat{\mathbf{z}}_{k|k-1}^{(\iota,i)}\not\in\mathcal{E}_{\mathbf{R}^{(j)}}^{\gamma_{\alpha_\chi}}\left(\mathbf{z}_k^{(i)}\right)\mid l=0\right) + \sum_{m=1}^L p_m 
    \end{split}
    \label{eq:outlier-prob-special}
\end{equation}
Where $p_m$ can be computed applying Assumption \ref{assum:known-natur-outlier-prob}. Subsequent derivations focus on the nominal case $l=0$.

\subsection{Probability of Outlier given a Posterior Distribution}

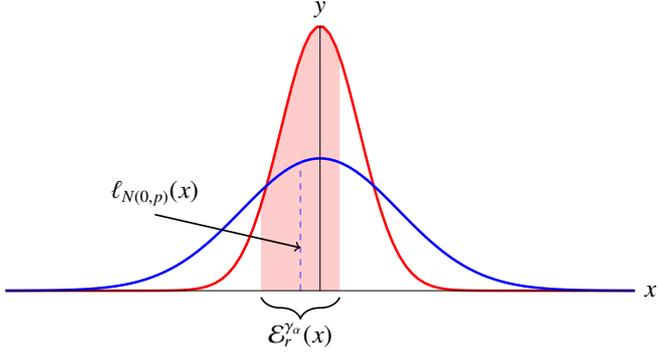
\begin{figure}[t]
    \centering
    \resizebox{\columnwidth}{!}{
        \input{Figures/GraphicalProof}}
    \caption{Graphical visualization of the proof of the outlier probability. The red curve is the measurement density function and the blue curve is the predicted density function.}
    \label{fig:unlikely_meas_graph_proof}
\end{figure}

The sequence $\delta_k^{(\iota,i)}$ is accumulated over a sliding window of size $W$. The windowed count follows a Poisson–Binomial distribution, since the Bernoulli trials have non-identical probabilities. Efficient \ac{cdf} computation methods exist \cite{fernandez_closed-form_2010,hong_computing_2013}.
\begin{prop}
    Let $\hat{\mathbf{z}}\!\sim\!N(\boldsymbol{\mu}_{\hat{z}},\mathbf{P})$ and $\mathbf{z}\!\sim\!N(\boldsymbol{\mu}_z,\mathbf{R})$ with $\mathbb{E}(\boldsymbol{\mu}_{\hat{z}})=\mathbb{E}(\boldsymbol{\mu}_z)$. For the region of inliers $\mathcal{E}_{\mathbf{R}}^{\gamma_{\alpha_\chi}}\!\left(\mathbf{z}\right)$, the probability of an inlier is
    \begin{equation}
        \begin{split}
            \bar{P}_\mathrm{in} &=   \mathbb{P}\!\left(\hat{\mathbf{z}}\in\mathcal{E}_{\mathbf{R}}^{\gamma_{\alpha_\chi}}\!\left(\mathbf{z}\right)\right) \\
            &=\int_{\mathbb{R}^{n_z}} \ell_{N\left(\mathbf{0},\mathbf{P} \right)}(\mathbf{x})\int_{\mathcal{E}_{\mathbf{R}}^{\gamma_{\alpha_\chi}}\left(\mathbf{x}\right)}\ell_{N\left(\mathbf{0},\mathbf{R}\right)}(\mathbf{y})\mathrm{d}\mathbf{y}\mathrm{d}\mathbf{x}
        \end{split}
        \label{eq:average_likely_prob}
    \end{equation}
    where $n_{z} = \mathrm{dim}\left(\mathbf{z}\right)$.
\end{prop}
\begin{proof}
    Using the law of total probability and the assumption that the Gaussian means coincide, such that the difference is zero (set to zero without loss of generality)
    \begin{equation}
        \begin{split}
            \bar{P}_{\mathrm{in}} &= \mathbb{P}\left(\hat{\mathbf{z}}\in\mathcal{E}_{\mathbf{R}}^{\gamma_{\alpha_\chi}}\left(\mathbf{z}\right)\right)\\
            &=\int_{\mathbb{R}^{n_z}}\ell_{N\left(\mathbf{0},\mathbf{P}\right)}(\mathbf{x})\mathbb{P}\left\{\hat{\mathbf{z}} \in \mathcal{E}_{\mathbf{R}}^{\gamma_{\alpha_\chi}}\left(\mathbf{z}\right)\mid\hat{\mathbf{z}}=\mathbf {x}\right\}{d}\mathbf{x} \\
            &= \int_{\mathbb{R}^{n_z}}\ell_{N\left(\mathbf{0},\mathbf{P}\right)}(\mathbf{x})\int_{\mathcal{E}_{\mathbf{R}}^{\gamma_{\alpha_\chi}}\left(\mathbf{x}\right)}\ell_{N\left(\mathbf{0},\mathbf{R}\right)}(\mathbf{y})\mathrm{d}\mathbf{y}\mathrm{d}\mathbf{x}
            \end{split}
    \end{equation}
\end{proof}
For the scalar case, where $\hat{z}\!\sim\!N(0,p)$ and $z\!\sim\!N(0,r)$, the inlier probability depends on the relative uncertainties $p$ and $r$. The likelihood $\ell_{N(0,p)}(y)$ evaluated at a specific point is illustrated in Fig.~\ref{fig:unlikely_meas_graph_proof}. For any $x\in\mathbb{R}$,
\begin{equation}
    \begin{split}
        &\ell_{N\left(0,p\right)}(x)\mathbb{P}\left(\hat{z}\in\mathcal{E}_{r}^{\gamma_{\alpha_\chi}}\left(z\right)\mid\hat{z}=x\right) \\
        =&\ell_{N\left(0,p\right)}(x) \mathbb{P}\left(z-\gamma_{\alpha_\chi}r \leq x\leq 
        z+\gamma_{\alpha_\chi}r\right)\\
        =&\ell_{N\left(\mathbf{0},p\right)}(x) \mathbb{P}\left(x-\gamma_{\alpha_\chi}r \leq z\leq x+\gamma_{\alpha_\chi}r\right)\\
        =& \ell_{N\left(\mathbf{0},p\right)}(x) \int_{x-\gamma_{\alpha_\chi}r}^{x+\gamma_{\alpha_\chi}r}\ell_{N\left(0,r\right)}(y)\mathrm{d}y
    \end{split}
\end{equation}Integrating over all $x$ yields Eq.~\eqref{eq:average_likely_prob}.  

Applying Eq.~\eqref{eq:predicted_non_infla} yields the system-specific form:
\begin{equation}
    \bar{P}^{(\iota,i)}_{\mathrm{in},k} = \int_{\mathbb{R}^{n_z^{(i)}}} \ell_{N\left(\mathbf{0},\mathbf{P}_{k|k-1}^{(\iota,i)} \right)}(\mathbf{x})\int_{\mathcal{E}_{\mathbf{R}^{(i)}}^{\gamma_{\alpha_\chi}}\left(\mathbf{x}\right)}\ell_{N\left(\mathbf{0},\mathbf{R}^{(i)}\right)}(\mathbf{y})\mathrm{d}\mathbf{y}\mathrm{d}\mathbf{x}
    \label{eq:average_likely_prob_sys}
\end{equation}

\subsubsection{Complexity of Likelihood Evaluation of Outlier Given Prior Distribution}

Evaluating Eq.~\eqref{eq:average_likely_prob_sys} involves nested integrals in $\mathbb{R}^{n_z^{(i)}}$, which is computationally intensive for $n_z^{(i)}\ge2$. For $n_z^{(i)}=1$, the integral can be expressed via the Gaussian error function. Marginalizing $\mathbf{P}_{k|k-1}^{(\iota,i)}$ and $\mathbf{R}^{(i)}$ yields $n_z^{(i)}$ univariate Gaussians with variances $\mathbf{P}_{k|k-1,q}^{(\iota,i)}$ and $\mathbf{R}_q^{(i)}$. The univariate inlier region is
\begin{equation}
    \mathcal{E}^{\gamma_{\alpha_\chi}}_{\mathbf{R}_q^{(i)}}(y) = \left\{x \; \middle| \; |x| \le y+\gamma_{\alpha_\chi}\mathbf{R}_q^{(i)}\right\}
    \label{eq:region_rect_univariate}
\end{equation}
and the average inlier probability simplifies to
\begin{equation}
    \bar{P}_{\mathrm{in},k,q}^{(\iota,i)} = \mathrm{erf}\left(\frac{\gamma_{\alpha_\chi}\mathbf{R}_q^{(i)}}{\sqrt{2\left(\left(\mathbf{R}_q^{(i)}\right)^2+\left(\mathbf{P}_{k|k-1,q}^{(\iota,i)}\right)^2\right)}}\right)
    \label{eq:average_likely_probsys_1dim-simple-eval}
\end{equation}
where $\mathrm{erf}(\cdot)$ is the Gaussian error function. See \ref{app:1} for how the Gaussian error function expression is derived.

\subsection{Windowed Count as Detection Mechanism}
\label{sec:binom_process_detect}

The complement of the Bernoulli variable is accumulated over a sliding window
\begin{equation}
        \Delta_{k,q}^{(\iota,i)} =\sum_{\kappa=k-W}^{k} \left(\delta^c\right)_{\kappa,q}^{(\iota,i)} 
        \label{eq:binom_dist}
\end{equation}
The maximum allowable count of outliers is obtained using the inverse Poisson–Binomial \ac{cdf}
\begin{equation}
    o_{\beta,q}^{(\iota,i)} = F_{\mathrm{PB}}^{-1}\left(\beta; W, \left\{P_{\mathrm{out},k,q}^{(\iota,i)}\right\}_{k=1}^{W}\right)
\end{equation}
where $\beta$ is a user-defined percentile. An alarm is raised when
\begin{equation}
    \lambda_k^{(\iota)} = \begin{cases}
    1 & \text{if }\Delta_{k,q}^{(\iota,i)}  > o_{\beta,q}^{(\iota,i)} ,\;\forall i,q\\
    0 & \text{otherwise}
    \end{cases}
\end{equation}
The sequences $(\delta^c)_{\kappa,q}^{(\iota,i)}$ and $\bar{P}_{\mathrm{in},\kappa,q}^{(\iota,i)}$ are stored in $(\mathcal{D}^c)_k^{(\iota)}$ and $\mathcal{P}_k^{(\iota)}$ for subsequent decision logic.

\section{Trajectory Stabilization Controller}
\label{sec:traj-controller}

The robot employs a nonlinear model predictive controller (\ac{nmpc}) for trajectory tracking. Depending on the operating state, the controller follows either a nominal trajectory or a path-replanned trajectory. The nominal trajectory is a sequence of time-indexed poses $\mathbf{t}_{\mathrm{n}} = (\mathbf{q}_{\mathrm{n},0}, \dots, \mathbf{q}_{\mathrm{n},K})$, where $K$ denotes the final time step. The path-replanned trajectory, generated as part of the mitigation process, is detailed in Section~\ref{sec:malicious-anomaly-mitigation}.

The \ac{nmpc} problem is formulated as
\begin{equation}
    \begin{split}
        \min_{\mathbf{x}_\kappa,\mathbf{u}_\kappa} &\quad \sum_{\kappa=k}^{k+M_\mathrm{c}}||\mathbf{q}_\kappa - \mathbf{q}_\mathrm{\kappa,ref}||_{\mathbf{W}_g}^2 +||\mathbf{q}_{k+M_\mathrm{c}} - \mathbf{q}_\mathrm{k+M_\mathrm{c},ref}||_{\mathbf{W}_g}^2\\
        \mathrm{s.t.} &\quad \mathbf{x}_0 = \mathbf{x}_\mathrm{init} \\
        &\quad \mathbf{x}_{\kappa+1} = \mathbf{f}_\kappa(\mathbf{x}_\kappa,\mathbf{u}_\kappa) \quad \text{for } \kappa = k,\dots,k+M_\mathrm{c} \\
        &\quad\mathbf{q}_\kappa = \mathbf{H}_q\mathbf{x}_{\kappa} \quad \text{for } \kappa = k,\dots, k+M_\mathrm{c} \\
        &\quad |\mathbf{u}_\kappa| \leq \mathbf{u}_\mathrm{max} \quad \text{for } \kappa = k,\dots,k+M_\mathrm{c}
    \end{split}
    \label{eq:nmpc}
\end{equation}
where $M_\mathrm{c}$ is the prediction horizon, $\mathbf{H}_q$ selects the pose vector $\mathbf{q}_k$ from the full state, $\mathbf{W}_g$ and $\mathbf{W}_f$ are weighting matrices, and $\mathbf{u}_\mathrm{max}$ defines input constraints. The reference poses $\mathbf{q}_{\kappa,\mathrm{ref}}$ are taken from either $\mathbf{t}_{\mathrm{n}}$ or the path-replanned trajectory, and $\mathbf{x}_\mathrm{init}$ denotes the initial state.

During operation and diagnosis, the initial state $\mathbf{x}_\mathrm{init}$ is set to $\boldsymbol{\mu}_{k|k}^{(\mathrm{op})}$ from the operational hypothesis $h_k^{(\mathrm{op})}$, and the controller tracks the nominal trajectory $\mathbf{t}_{\mathrm{n}}$. Upon solving the optimization, the predicted control inputs $\mathbf{u}_{\mathrm{p},k,\kappa}^{(\mathrm{op})}$ are used to propagate all hypotheses $\mathcal{H}_k$, yielding predicted states $\boldsymbol{\mu}_{\mathrm{p},k,\kappa}^{(\iota)}$.

When the robot enters the Mitigation state, the state machine activates the path-replanner module, which generates a trajectory $\mathbf{t}_{\mathrm{rp},k}^{(\iota,j)}$. The \ac{nmpc} then uses this trajectory as reference, with the corresponding hypothesis $h_k^{(\iota)}$ providing the initial state. The predicted inputs obtained during mitigation are likewise used to propagate all remaining hypotheses and the operational hypothesis for consistency.

\section{Malicious Anomaly Mitigation}
\label{sec:malicious-anomaly-mitigation}

This section presents the mitigation strategy activated once a malicious anomaly is detected. At time step $k_m$, the robot maintains a set of hypotheses $\mathcal{H}_{k_m}$ with disjoint measurement source sets $\uplus^{(\iota)} \mathcal{O}_k^{(\iota)} = \mathcal{O}_k$. To discriminate among these hypotheses, the robot must deviate from the nominal trajectory to collect additional information, while minimizing performance degradation.

The robot navigates so that each hypothesis visits designated viewpoints, enabling new exteroceptive measurements for hypothesis validation. The binary variable $\delta_{\mathrm{ps},k}^{(\iota,i,j)}$, defined in Section~\ref{seq:seq:view-point-region}, indicates whether a viewpoint is visited by by $h^{(\iota)}$, used as basis for acceptance and rejection of hypotheses.

\subsection{Performance Loss Index}
The nominal trajectory $\mathbf{t}_\mathrm{n}$ satisfies mission objectives as long as the robot remains within a disc of radius $\rho$ centered at the nominal pose. Deviation from this region results in a gradual performance loss that increases with both distance and duration of violation. For hypothesis $\iota$, the violation distance is defined as
\begin{equation}
    d_{\rho,k}^{(\iota)} = \begin{cases}
        ||\mathbf{p}_{k}^{(\iota)}-\mathbf{p}_{n,k}|| - \rho & \text{if }  ||\mathbf{p}_{k}^{(\iota)}-\mathbf{p}_{n,k}|| - \rho > 0 \\
        0 & \text{otherwise}
    \end{cases}
\end{equation}
The accumulated violation time $v_k^{(\iota)}$ evolves as
\begin{equation}
    v_{k}^{(\iota)} = \begin{cases}
        \max\left(0,v_{k-1}^{(\iota)}-1\right) & \text{if } ||\mathbf{p}_{k}^{(\iota)}-\mathbf{p}_{n,k}|| - \rho \leq 0 \\
        v_{k-1}^{(\iota)} + 1 & \text{if } ||\mathbf{p}_{k}^{(\iota)}-\mathbf{p}_{n,k}|| - \rho > 0
    \end{cases} 
\end{equation}
The combined effect yields the performance loss index
\begin{equation}
    P_{\mathrm{loss},k}^{(\iota)} = 1-\mathrm{e}^{-\alpha_1d_{\rho,k}^{(\iota)}-\alpha_2v_{k}^{(\iota)}}
    \label{eq:performance-loss-index}
\end{equation}
where $\alpha_1$ and $\alpha_2$ are weighting factors related to the maximum allowable deviation $\mathrm{d}_{\max}$ and maximum time $v_{\max}$ outside the nominal region. The index is computed for all hypotheses $h_k^{(\iota)} \in \mathcal{H}_k$ and for $h_k^{(\mathrm{op})}$.

\subsection{Path Re-Planner}

At time step $k_m$, the robot maintains $|\mathcal{H}_{k_m}|$ hypotheses and can access $|\mathcal{VP}|$ viewpoints, yielding $|\mathcal{H}_{k_m}| \times |\mathcal{VP}|$ possible re-planned trajectories. Since the truthful hypothesis is unknown, all combinations must be considered. The selected trajectory $\mathbf{t}_{\mathrm{rp},k}^{(\iota,j)}$ for hypothesis $\iota$ and viewpoint $j$ should also minimize performance loss across alternative hypotheses $\nu$, ensuring minimal degradation if $\iota$ is later rejected.

\begin{figure}[t]
    \centering
    \includegraphics[width=\linewidth]{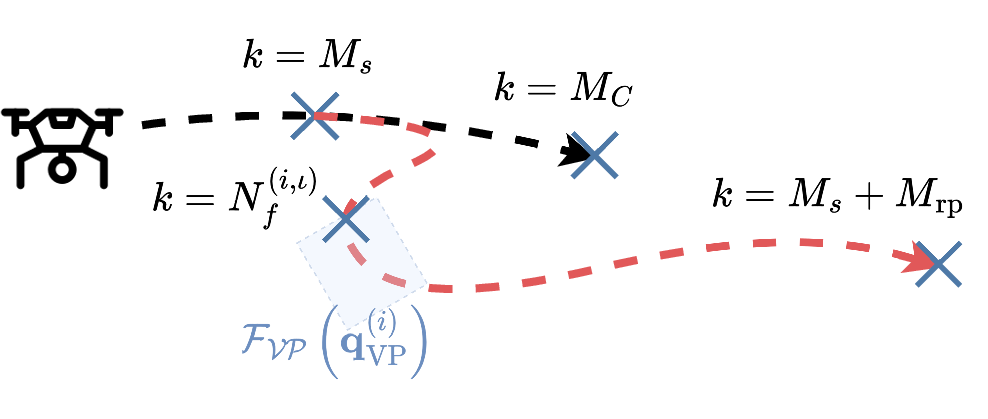}
    \caption{Illustration of how the path re-planned trajectory is computed using a predicted state from the \ac{nmpc} trajectory stabilization controller. The predicted states are shown in black and the path re-planned trajectory is shown in red. The blue crosses shows the time instances and the blue rectangle shows the view point region.}
    \label{fig:planning_example}
\end{figure}

Because the computation of all trajectories is intensive, re-planning is performed at a lower frequency than the \ac{nmpc}. The path-replanner horizon $M_\mathrm{rp}$ therefore exceeds the controller horizon $M_\mathrm{c}$ and is sufficiently large for meaningful loss evaluation. The re-planner uses the \ac{nmpc} predicted state at step $M_\mathrm{s} \le M_\mathrm{c}$ as its initial condition, as illustrated in Fig.~\ref{fig:planning_example}.

The re-planning exploits differential flatness \cite{mellinger_minimum_2011}, enabling trajectory generation as piecewise polynomial splines with reduced dimensionality. The flat outputs $\mathbf{z}(t)$, a subset of $\mathbf{x}(t)$, determine all remaining states and inputs algebraically. For floating-base platforms such as \ac{uav}, the system must be twice differentiable \cite{mellinger_minimum_2011}.

Each hypothesis $\iota$ must reach the vicinity of viewpoint $j$, i.e., $\mathbf{q}_{k|k}^{(\iota)}\in\mathcal{F}_{\mathcal{VP}}(\mathbf{q}_\mathrm{VP}^{(j)})$ for $\tau_{\mathrm{r/a}}$ time steps. Let $N_f^{(\iota,j)}$ denote the time at first entry into this region. Trajectories are modeled as fourth-degree splines with $C^2$ continuity. The re-planning problem minimizing deviation from both the viewpoint and the nominal path is formulated as

\begin{equation}
    \begin{split}
        \min_{\mathbf{P}_m^{(\iota,j)}} &\quad \sum_{\kappa=k}^{k+N_{f,\tau}^{(\iota,j)}}\left|\left|\mathbf{q}^{(\iota)}_\kappa-\mathbf{q}_\mathrm{VP}^{(j)}\right|\right|_{\mathbf{W}_{\mathrm{VP}}}^2+\sum_{\kappa=k+N_{f,\tau}^{(\iota,j)}}^{M_\mathrm{rp}} \left|\left|\mathbf{q}_\kappa^{(\iota)}-\mathbf{q}_{\mathrm{n},\kappa}\right|\right|_{\mathbf{W}_g}^2  \\
        &\quad\quad + \left|\left|\mathbf{q}_{k+M_\mathrm{rp}}^{(\iota)}+\mathbf{q}_{\mathrm{n},k+M_\mathrm{rp}}\right|\right|_{\mathbf{W}_f}^2 \\
        \mathrm{s.t.} &\quad  m = \left\lfloor\frac{\kappa}{4}\right\rfloor, \;\; |\ddot{\mathbf{q}}_\kappa^{(\iota)}| \leq \ddot{\mathbf{q}}_\mathrm{max} \\
        &\quad \dot{\mathbf{q}}_0^{(\iota)} = \dot{\mathbf{q}}_\mathrm{init}, \;\; \mathbf{q}_0^{(\iota)} = \mathbf{q}_\mathrm{init} \\
        &\quad\mathbf{q}_\kappa^{(\iota)} = \mathbf{P}_m^{(\iota,j)}\left(\kappa T_s\right) \quad \text{for } \kappa = k,\dots, k+M_\mathrm{rp} \\
        &\quad \ddot{\mathbf{q}}_\kappa^{(\iota)} = \ddot{\mathbf{P}}_m^{(\iota,j)}(\kappa T_s) \quad \text{for } \kappa = k,\dots, k+M_\mathrm{rp} \\ 
        &\quad \mathbf{q}_\kappa^{(\iota)} \in \mathcal{F}\left(\mathbf{q}_\mathrm{VP}^{(j)}\right) \quad \text{for } \kappa=k+N_f^{(\iota,j)},\dots,k+N_{f,\tau}^{(\iota,j)}\\
        &\quad \mathbf{P}_{m}^{(\iota,j)}(T_s(4m+3)) = \mathbf{P}_{m+1}^{(\iota,j)}(4T_sm) \quad \\
        &\quad\quad\quad\quad\quad\quad\quad\quad\;\;\text{for } m=0,\dots,M_{\mathrm{rp}}/4\\
    \end{split}
    \label{eq:differential-flatness-program}
\end{equation}
where $\mathbf{P}_m^{(\iota,j)}$ are the piecewise polynomial functions of the spline defined on the interval $[4T_sm,T_s(4m+3)]$, $\ddot{\mathbf{z}}_\mathrm{max}$ is a constraint on the acceleration that relates to the input constraints, $\mathbf{W}_\mathrm{VP}$, $\mathbf{W}_g$ and $\mathbf{W}_f$ are tuning matrices and $N_{f,\tau}^{(\iota,j)} = N_f^{(\iota,j)} + \tau_{\mathrm{r/a}}$. $\mathbf{q}_\mathrm{init}$ and $\dot{\mathbf{q}}_\mathrm{init}$ are pose and velocities from the $M_s$ time step in the \ac{nmpc} prediction. The program provides coefficients for the polynomials $\mathbf{P}_m^{(\iota,j)}$, from which we determine $\mathbf{t}_{\mathrm{rp},k}^{(\iota,j)}$.

Alternative trajectories for the remaining hypotheses $\nu$ are generated for $\kappa \ge N_{f,\tau}^{(\iota,j)}$ to account for their expected performance loss if $\iota$ is rejected. The corresponding optimization, analogous to Eq.~\eqref{eq:differential-flatness-program}, is given in Eq.~\eqref{eq:differential-flatness-program-alternative}
\begin{equation}
    \begin{split}
        \min_{\mathbf{P}_m^{(\iota,\nu,j)}} &\quad \sum_{\kappa=k+N_{f,\tau}^{(\iota,j)}}^{M_\mathrm{rp}} ||\mathbf{q}_\kappa^{(\nu)}-\mathbf{q}_{\mathrm{n},\kappa}||_{\mathbf{W}_g}^2 \\
        \mathrm{s.t.} & \quad  m = \left\lfloor\frac{\kappa}{4}\right\rfloor, \;\; |\ddot{\mathbf{q}}_\kappa^{(\nu)}| \leq \ddot{\mathbf{q}}_\mathrm{max}\\
        &\quad \mathbf{q}_{k+N_{f,\tau}}^{(\nu)} = \mathbf{q}_{k+N_{f,\tau}}^{(\iota,j)}, \;\; \dot{\mathbf{q}}_{k+N_{f,\tau}}^{(\nu)} = \dot{\mathbf{q}}_{k+N_{f,\tau}}^{(\iota,j)} \\
        &\quad \mathbf{q}_\kappa^{(\nu)} = \mathbf{P}_m^{(\nu)}\left(\kappa T_s\right) \\
        &\quad\quad\quad \text{for } \kappa = k+N_{f,\tau}^{(\iota,j)},\dots, k+M_\mathrm{rp} \\
        &\quad \ddot{\mathbf{q}}_\kappa^{(\nu)} = \ddot{\mathbf{P}}_m^{(\iota,\nu,j)}(\kappa T_s) \\
        &\quad\quad\quad \text{for } \kappa = k+N_{f,\tau}^{(\iota,j)},\dots, k+M_\mathrm{rp} \\
        &\quad \mathbf{P}_{m}^{(\iota,\nu,j)}(T_s(4m+3)) = \mathbf{P}_{m+1}^{(\iota,\nu,j)}(4T_sm) \quad \\
        &\quad\quad\quad\text{for } m=N_{f,\tau}^{(\iota,j)}/4,\dots,M_{\mathrm{rp}}/4
    \end{split}
    \label{eq:differential-flatness-program-alternative}
\end{equation}
The full problem could be formulated as a \ac{mip} by introducing binary variables for viewpoint entry and performance loss, but this is computationally intractable due to the relatively long prediction window $M_\mathrm{rp}$. Instead, a suboptimal heuristic based on ternary search is proposed, iteratively solving quadratic programs to minimize $N_f^{(\iota,j)}$ and indirectly reduce performance loss, as shown in Algorithm~\ref{alg:ternary-search}.

\begin{algorithm}
    \caption{Ternary search for minimizing Eq. \ref{eq:differential-flatness-program} with respect to $N_{f,k}^{(\iota,j)}$}\label{alg:ternary-search}
    \KwData{$\mathbf{q}_0^{(\iota)}$, $\dot{\mathbf{q}}_0^{(\iota)}$, $v_{k}^{(\iota)}$, $\mathbf{q}_{\mathrm{VP}}^{(j)}$, $\mathbf{W}_\mathrm{VP}$, $\mathbf{W}_g$}
    \KwResult{$\mathbf{t}_{\mathrm{rp},k}^{(\iota,j)}$, $\mathbf{u}_{\mathrm{rp},k}^{(\iota,j)}$, $N_{f,k}^{(\iota,j)}$, $f_{\mathrm{obj}}^{(\iota,j)}$}
    $m_{\mathrm{min}} \gets 0, \;m_{\mathrm{max}} \gets M_\mathrm{rp}$\;
    $N_{f,1} \gets m_{\min} + \left\lfloor\frac{m_{\mathrm{max}}-m_{\mathrm{min}}}{3}\right\rceil$\;
    $N_{f,2} \gets m_{\mathrm{max}} - \left\lfloor\frac{m_{\mathrm{max}}-m_{\mathrm{min}}}{3}\right\rceil$\;
    \While{$m_1 \leq m_2$}{
        $N_{f,1} \gets m_{\min} + \left\lfloor\frac{m_{\mathrm{max}}-m_{\mathrm{min}}}{3}\right\rceil$\;
        $N_{f,2} \gets m_{\mathrm{max}} - \left\lfloor\frac{m_{\mathrm{max}}-m_{\mathrm{min}}}{3}\right\rceil$\;
        $\mathbf{u}_{\mathrm{rp},k,1}, \mathbf{t}_{\mathrm{rp},k,1}, \mathbf{u}_{\mathrm{rp},k,2}, \mathbf{t}_{\mathrm{rp},k,2} \gets $ solution from Eq.~\eqref{eq:differential-flatness-program} using $N_{f,1}$ and $N_{f,2}$ respectively\;
        Compute $P_{\mathrm{loss},k,1}$ and $P_{\mathrm{loss},k,2}$ according to Eq. \eqref{eq:performance-loss-index}\;
        \If{Eq. \eqref{eq:differential-flatness-program} using $N_{f,1}$ or $N_{f,2}$ is not feasible}{
            Set $m_\mathrm{min}$ to either $N_{f,1} + 1$ or $N_{f,2}+1$\;
            continue\;
        }
        \eIf{$P_{\mathrm{loss},1}> P_{\mathrm{loss},2}$}{
            $m_{\mathrm{min}} \gets N_{f,1}$\;
        }{
            $m_\mathrm{max} \gets N_{f,2}$\;
        }
    }
    Return $\mathbf{t}_{\mathrm{rp},k,1}, \;\mathbf{u}_{\mathrm{rp},k,1}, \; N_{f,1}, \;\max_{0,\leq k\leq M_{\mathrm{rp}}} P_{\mathrm{loss},k,1}$\;
\end{algorithm}
Finally, Algorithm~\ref{alg:optimal-re-plan} summarizes the overall computation of the optimal re-plan across all hypotheses and viewpoints. The selected pair $(\iota_\mathrm{sel}, j_\mathrm{sel})$ minimizes the cumulative performance loss.

\begin{algorithm}
    \caption{Computation of optimal re-plan}
    \label{alg:optimal-re-plan}
    \KwData{$\mathcal{H}_k$, $\mathcal{VP}$, $\mathbf{t}_\mathrm{n}$, $\mathbf{W}_\mathrm{VP}$, $\mathbf{W}_g$, $\left\{v_k^{(\iota)}\right\}_{\iota}^{|\mathcal{H}_k|}$}
    \KwResult{$\iota_\mathrm{sel}$, $j_\mathrm{sel}$, $\mathbf{t}_{\mathrm{rp},k}^{(\iota_\mathrm{sel},j_\mathrm{sel})}$, $\mathbf{t}_{\mathrm{rp},k}^{(\iota_\mathrm{sel},\nu,j_\mathrm{sel})}$}
    \For{$h_k^{(\iota)}\in\mathcal{H}_k$}{
        \For{$\mathbf{q}_\mathrm{VP}^{(j)} \in \mathcal{VP}$}{
            Solve Algorithm~\ref{alg:ternary-search} for $(\iota,j)$;\\
            For each $\nu\neq\iota$, propagate $\mathbf{t}_{\mathrm{rp}}^{(\iota,\nu,j)}$ via Eq.~\eqref{eq:differential-flatness-program-alternative};\\
            Compute $P_{\mathrm{loss}}^{(\iota,\nu,j)}$ and accumulate total loss;\\
        }
    }
    Select $(\iota_\mathrm{sel}, j_\mathrm{sel}) = \arg\min f_{\mathrm{obj}}$.
\end{algorithm}

\section{Case Studies}
\label{sec:case-studies}
This section evaluates the proposed algorithm through two case studies. The first analyzes the sensitivity of the detection mechanism to bias attacks, while the second demonstrates the full algorithm in a complete scenario.

The kinematic motion model used throughout the case studies is
\begin{equation}
    \begin{split}
        \mathbf{x}_{k+1} &= \mathbf{x}_k + T_s
        \begin{bmatrix}
            \left(\mathbf{R}\left(\theta_k\right)\mathbf{v}_k\right)^\mathrm{T} & a_{x,k} & a_{y,k} & \omega_k
        \end{bmatrix}^\mathrm{T} \\
        \mathbf{R}(\theta) &= 
        \begin{bmatrix} 
            \cos(\theta) & -\sin(\theta) \\ \sin(\theta) &  \cos(\theta)
        \end{bmatrix} \\
        \mathbf{v}_k &= 
        \begin{bmatrix}
            v_{x,k} & v_{y,k}
        \end{bmatrix}^{\mathrm{T}}
    \end{split}
    \label{eq:scenario_motion_model}
\end{equation}
The state vector is $\mathbf{x} = \begin{bmatrix}\mathbf{p}^\mathrm{T}& \mathbf{v}^\mathrm{T}& \theta
\end{bmatrix}^\mathrm{T}$, with $\mathbf{p} = \begin{bmatrix}x & y\end{bmatrix}^\mathrm{T}$ and $\mathbf{v} = \begin{bmatrix}v_x & v_y
\end{bmatrix}^\mathrm{T}$. The input vector is $\mathbf{u} = \begin{bmatrix}a_x & a_y & \omega\end{bmatrix}^\mathrm{T}$, and the sampling time $T_s = 0.1\,\mathrm{s}$. The process noise covariance matrix is $\mathbf{Q} = \texttt{diag}\left(\begin{bmatrix}\sigma_x^2 & \sigma_y^2 & \sigma_{v_x}^2 & \sigma_{v_y}^2 & \sigma_\theta^2\end{bmatrix}\right)$ with $\sigma_x = \sigma_y=0.71\mathrm{m}$, $\sigma_{v_x}=\sigma_{v_y}=0.01\frac{\mathrm{m}}{\mathrm{s}}$ and $\sigma_\theta=2.38^\circ$. The \ac{imu} covariance matrix is $\mathbf{Q}_\mathrm{IMU} = \texttt{diag}\left(\begin{bmatrix} \sigma_{a_x}^2 & \sigma_{a_y}^2 & \sigma_\omega^2 \end{bmatrix}\right)$ where $\sigma_{a_x}=\sigma_{a_y}=3.16\cdot10^{-2}\frac{\mathrm{m}}{\mathrm{s}^2}$ and $\sigma_\omega=4.47\cdot10^{-3}\frac{\mathrm{rad}}{\mathrm{s}}$

The \ac{rf} sensor provides range, \ac{aoa}, and \ac{aod} relative to the $j$th \ac{rf} anchor according to
\begin{equation}
    \label{eq:meas}
    \mathbf{z}_k^{(i)} = \mathbf{h}_{\mathrm{RF}}\left(\mathbf{q}_k^{(l)}, \mathbf{p}_{\mathrm{RF}}^{(i)},\boldsymbol{\epsilon}_k\right) + \mathbf{w}_{\mathrm{RF},k}^{(j,l)}
\end{equation}
where
\begin{align}
    \mathbf{z}_k^{(i)} &=
    \begin{bmatrix}
        r_k^{(i)} & \theta_{\mathrm{AOA},k}^{(i)} & \theta_{\mathrm{AOD},k}^{(i)}
    \end{bmatrix}^\mathrm{T} \\
    &\mathbf{h}\left(\mathbf{q}_k,\mathbf{p}_{\mathrm{RF},k}^{(i)},\boldsymbol{\epsilon}_k^{(i)}\right)= \\
    &
    \begin{bmatrix}
        ||\mathbf{p}_{\mathrm{RF},k}^{(i)} - \mathbf{p}_k + \boldsymbol{\epsilon}_k|| \\ \pi +\mathrm{atan2}\left(y_k^{(i)} - y_k + \epsilon_{y,k}, x_k^{(i)} - x_k + \epsilon_{x,k}\right) -\theta_k \\
        \mathrm{atan2}\left(y_k^{(i)} - y_k + \epsilon_{y,k}, x_k^{(i)} - x_k +\epsilon_{x,k}\right) -\theta_k
    \end{bmatrix}
\end{align}
There are two modes, $l=0,1$. The second mode occurs with $10\%$ probability and biases the $x$-direction with $5\mathrm{m}$. The same noise parameters are used for both modes $\mathbf{w}_{\mathrm{RF},k}^{(i)} \sim N(\mathbf{0}, \mathbf{R}^{(i)}_\mathrm{RF})$ with $\mathbf{R}^{(i)}_\mathrm{RF} = \texttt{diag}([\sigma_r^2, \sigma_{\theta_\mathrm{AOA}}^2, \sigma_{\theta_\mathrm{AOD}}^2])$, $\sigma_r = 1\,\mathrm{m}$, and $\sigma_{\theta_\mathrm{AOA}} = \sigma_{\theta_\mathrm{AOD}} = 0.5^\circ$. The communication range of the \ac{rf} device is $100\,\mathrm{m}$. The \ac{gnss} model is
\begin{equation}
    \begin{split}
        \mathbf{z}_{\mathrm{GNSS}}  &= \mathbf{h}_\mathrm{GNSS}\left(\mathbf{p}_k,\boldsymbol{\epsilon}_{\mathrm{GNSS},k}\right) + \mathbf{w}_{\mathrm{GNSS},k} \\
        &= \mathbf{p}_k + \boldsymbol{\epsilon}_{\mathrm{GNSS},k} + \mathbf{w}_{\mathrm{GNSS},k}
    \end{split}
\end{equation}
where $\mathbf{w}_{\mathrm{GNSS},k} \sim N\left(\mathbf{0}, \mathbf{R}_\mathrm{GNSS}\right)$, $\mathbf{R}_\mathrm{GNSS} = \texttt{diag}\left(\begin{bmatrix}\sigma_{\mathrm{G,}x}^2 & \sigma_{\mathrm{G},y}^2 \end{bmatrix}\right)$, and $\sigma_{\mathrm{G},x} = 1\mathrm{m}$ and $\sigma_{\mathrm{G},y} = 1\mathrm{m}$. 

The adversarial signal $\boldsymbol{\epsilon}_k = \begin{bmatrix}\epsilon_{x,k}& \epsilon_{y,k}\end{bmatrix}^\mathrm{T}$ models the spoofing signal. All measurements are collected synchronously.

The parameters $\alpha_1 = 1.73\cdot10^{-2}$ and $\alpha_2 =3.35\cdot10^{(-3}$ in Eq. \eqref{eq:performance-loss-index} are determined by allowing for a $50\%$ performance loss, a maximum deviation of $50\mathrm{m}$ and $200\mathrm{s}$ outside of the disc respectively. The windowed count is $W=50$ and $W_\mathrm{p} = 5$.

\subsection{Analyzing the Sensitivity of the Algorithm to Bias Attacks}

The first case study evaluates the algorithm’s sensitivity to biasing cyber-attacks through the windowed count mechanism and the state transitions of the diagnosis module. Synthetic data are generated for combinations of $\alpha_\chi$, $\beta$, and $\alpha_\mathrm{F}$, each repeated over 200 realizations to estimate the \ac{fpr} and \ac{tpr}. Only constant biases in the $x$-direction are considered, active from initialization.

A false positive is defined as an erroneous transition to the diagnosis state, while a true positive denotes correct maintenance of the operational state. For mitigation evaluation, a true positive indicates correct identification of hypotheses $\mathcal{O}_k^{(\iota)}$ when an attack is present, and a true negative corresponds to returning to the operation state in its absence.

\begin{figure}[t]
    \centering
    \includegraphics[width=0.6\linewidth]{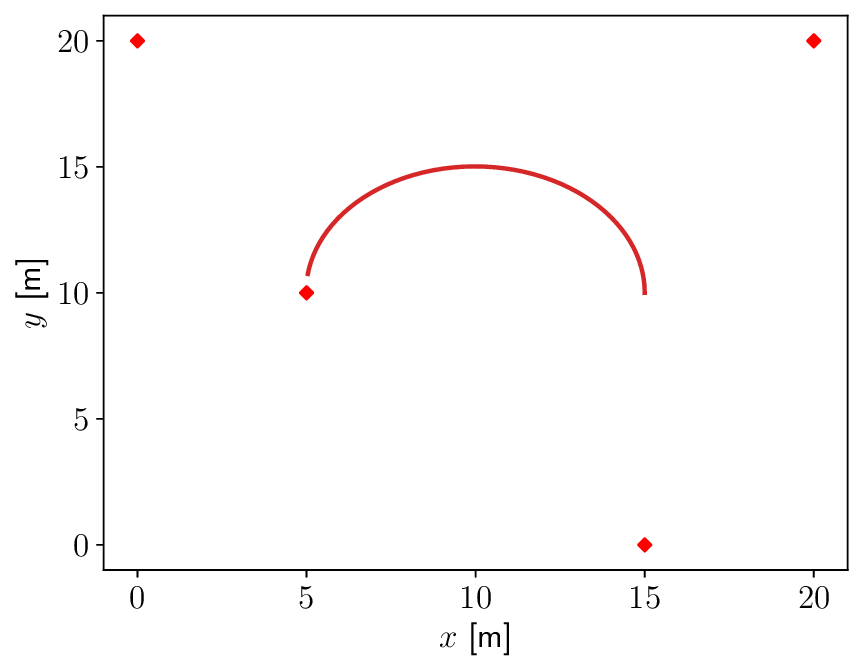}
    \caption{Scenario for sensitivity analysis. The robot follows a circular trajectory (red) and communicates with four \ac{rf} anchors (red diamonds).}
    \label{fig:sens:scenario1}
\end{figure}

The setup consists of one \ac{gnss} receiver and four \ac{rf} anchors, as shown in Fig.~\ref{fig:sens:scenario1}.

\begin{figure}[t]
    \centering
    \subfloat[\label{fig:sens:scenario1_resFPR_COUNTING}]{
        \includegraphics[width=0.70\linewidth]{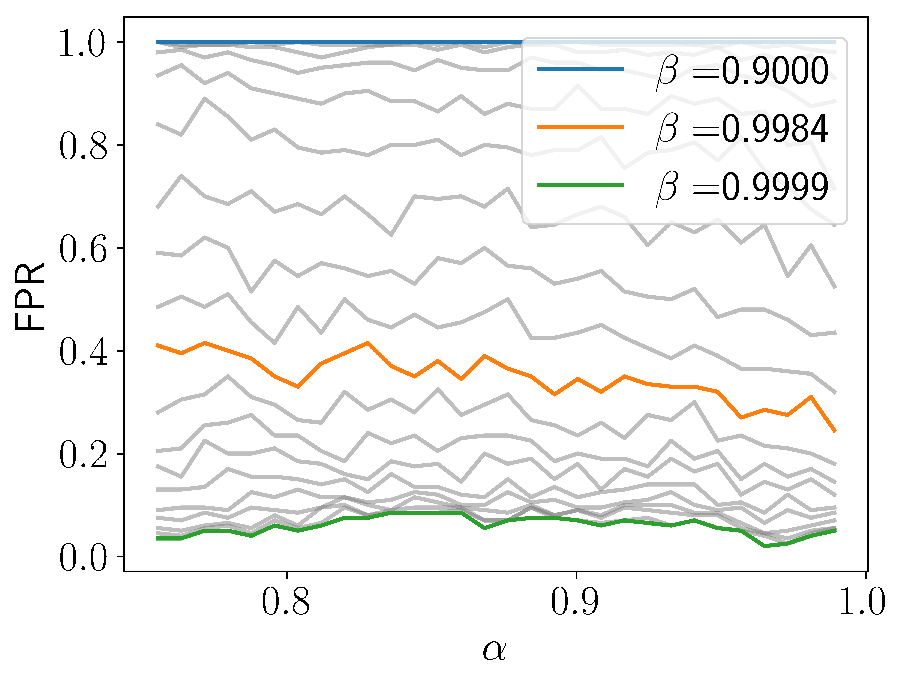}
    }\\
    \subfloat[\label{fig:sens:scenario1_resFPR_SPLIT}]{
        \includegraphics[width=0.70\linewidth]{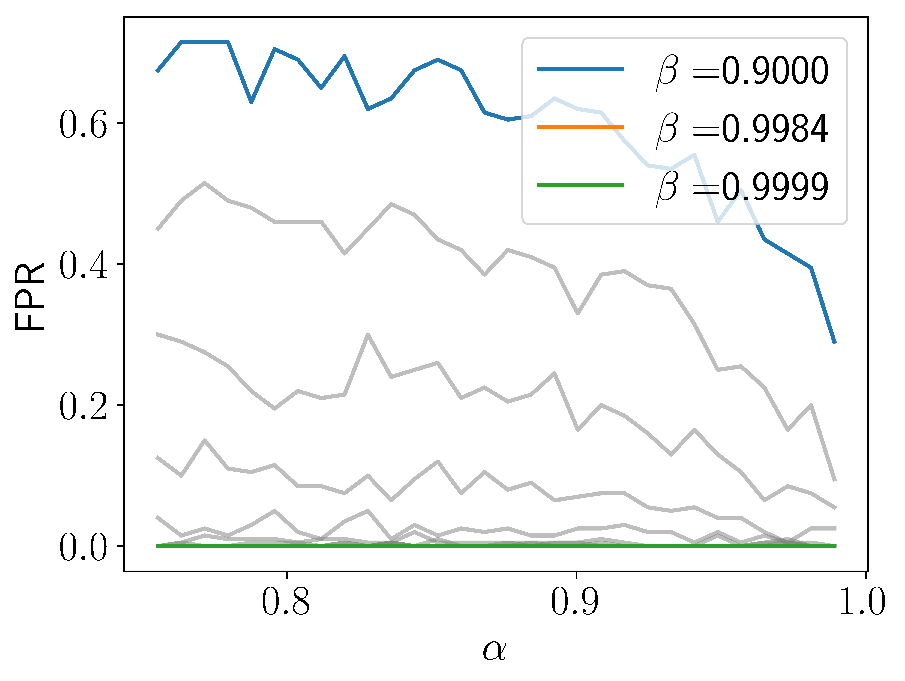}
    }
    \caption{(a) shows the \acl{fpr} of the algorithm when the robot is supposed to stay in the operation state and in (b) move from the diagnosis state to the operation state, recovering a single hypothesis. Both are a function of outlier percentile $\alpha_\chi$. In (b) the robot is initialized in the diagnosis state with $|\mathcal{O}_k^{(\iota)}|=4\; \forall \iota$. In (b), the green curve is ontop of the orange. The gray curves are other Poisson binomial percentiles $\beta$ in the range from $90.00\%$ to $99.99\%$, we highlight three.}
    \label{fig:sens:scenario1_resFPR}
\end{figure}

\begin{figure*}[t]
    \centering
    \subfloat[\label{fig:sens:scenario1_res:TPR90.00}]{
        \includegraphics[width=0.32\linewidth]{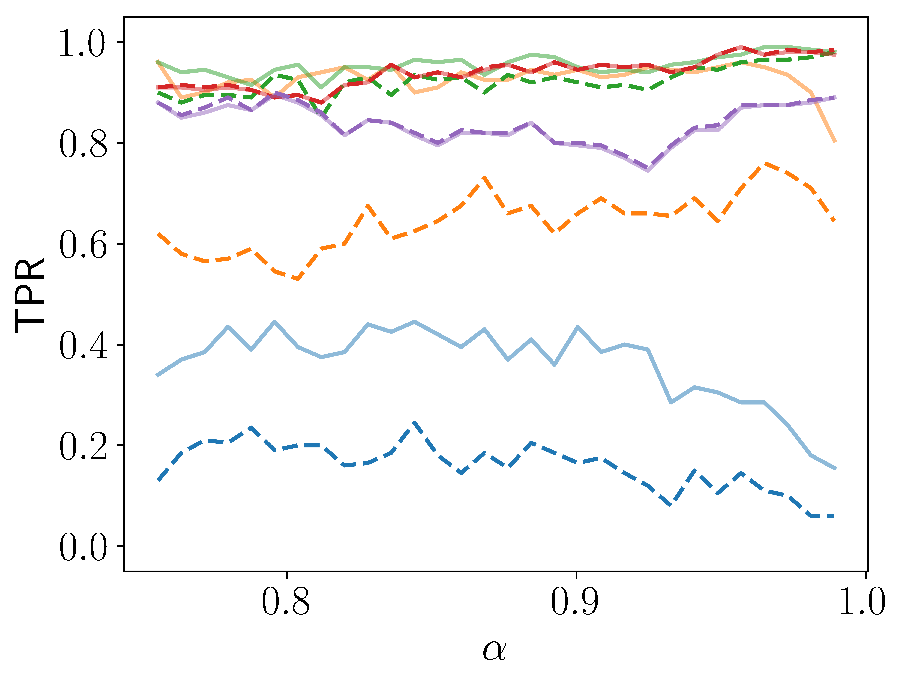}
    }
    \subfloat[\label{fig:sens:scenario1_res:TPR99.95}]{
        \includegraphics[width=0.32\linewidth]{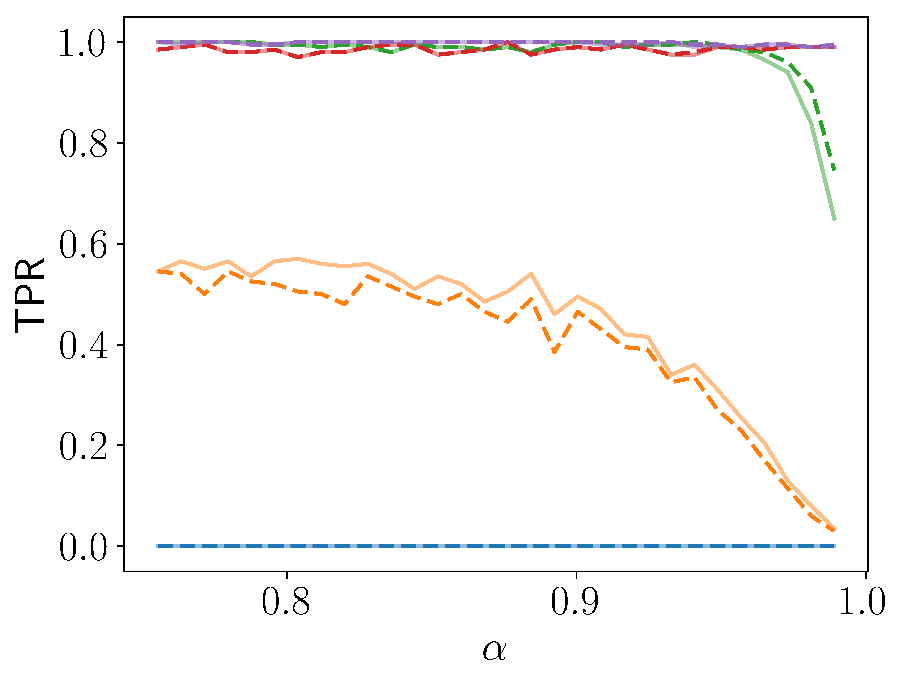}
    }
    \subfloat[\label{fig:sens:scenario1_res:TPR99.99}]{
        \includegraphics[width=0.32\linewidth]{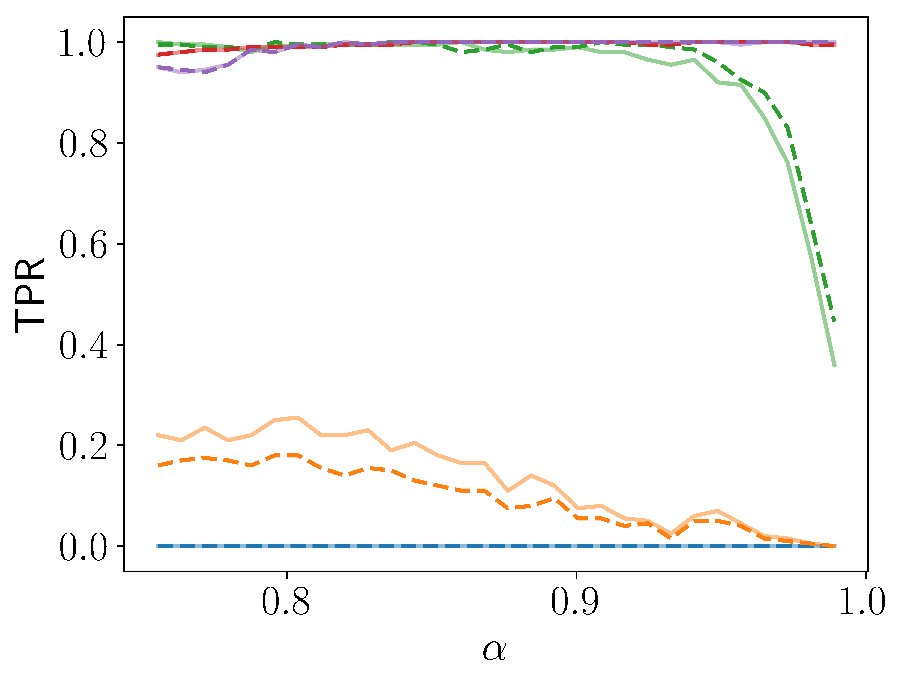}
    }
    \caption{
    True positive rate (\acl{tpr}) for $\beta = 90.00\%$ in (a), $\beta = 99.84\%$ in (b), and $\beta = 99.99\%$ in (c). The corresponding \ac{fpr} values are shown in Fig.~\ref{fig:sens:scenario1_resFPR}. The blue, orange, green, red and purple plots show the \ac{tpr} when the bias is of size 1m, 2m, 3m, 4m and 5m in the $x$-direction respectively. The low transparency solid lines and the dashed lines show the \ac{tpr} when $\alpha_\mathrm{F}$ is $38.29\%$ and $95.44\%$ respectively, spanning any value in between.}
    \label{fig:sens:scenario1_resTPR}
\end{figure*}

The resulting \ac{fpr} and \ac{tpr} are summarized in Figs.~\ref{fig:sens:scenario1_resFPR} and \ref{fig:sens:scenario1_resTPR}. High \ac{fpr} values occur at $\beta = 90\%$ but decrease as $\beta \rightarrow 99.99\%$. This reflects the effect of the Poisson binomial percentile on the confidence of the count process. The transition logic from diagnosis to operation, based on Eq.~\eqref{eq:sm-diagnosis-logic}, maintains low \ac{fpr} across all $\beta$ values due to consistent hypothesis merging when no attack is present.

Figure~\ref{fig:sens:scenario1_resTPR} shows the \ac{tpr} under increasing bias magnitudes. The small bias, $1\,\mathrm{m}$, remain below the outlier threshold $\alpha_\chi$, yielding low detection probability to non-existing, while larger biases, $\geq2\,\mathrm{m}$, increase detection likelihood for smaller $\alpha_\chi$. Higher $\beta$ values reduce both false positives and detection sensitivity. The threshold $\alpha_\mathrm{F}$ has diminishing influence as $\beta$ increase. 

At high Poisson binomial percentiles, as illustrated in Fig.~\ref{fig:sens:scenario1_res:TPR99.99}, the detection probability for $3\,\mathrm{m}$ biases declines at high $\alpha_\chi$. For these combinations of $\alpha_\chi$ and $\beta$, the allowable count of outliers represents only a small portion of the probability mass, rendering such deviations statistically rare, and subsequently the \ac{tpr} decreases. The parameter $\alpha_\mathrm{F}$ exerts diminishing influence at higher $\beta$ values, as the detection process becomes dominated by the outlier count threshold.

\subsection{Complete Algorithm}
\begin{figure}[t]
    \centering
    \includegraphics[width=\linewidth]{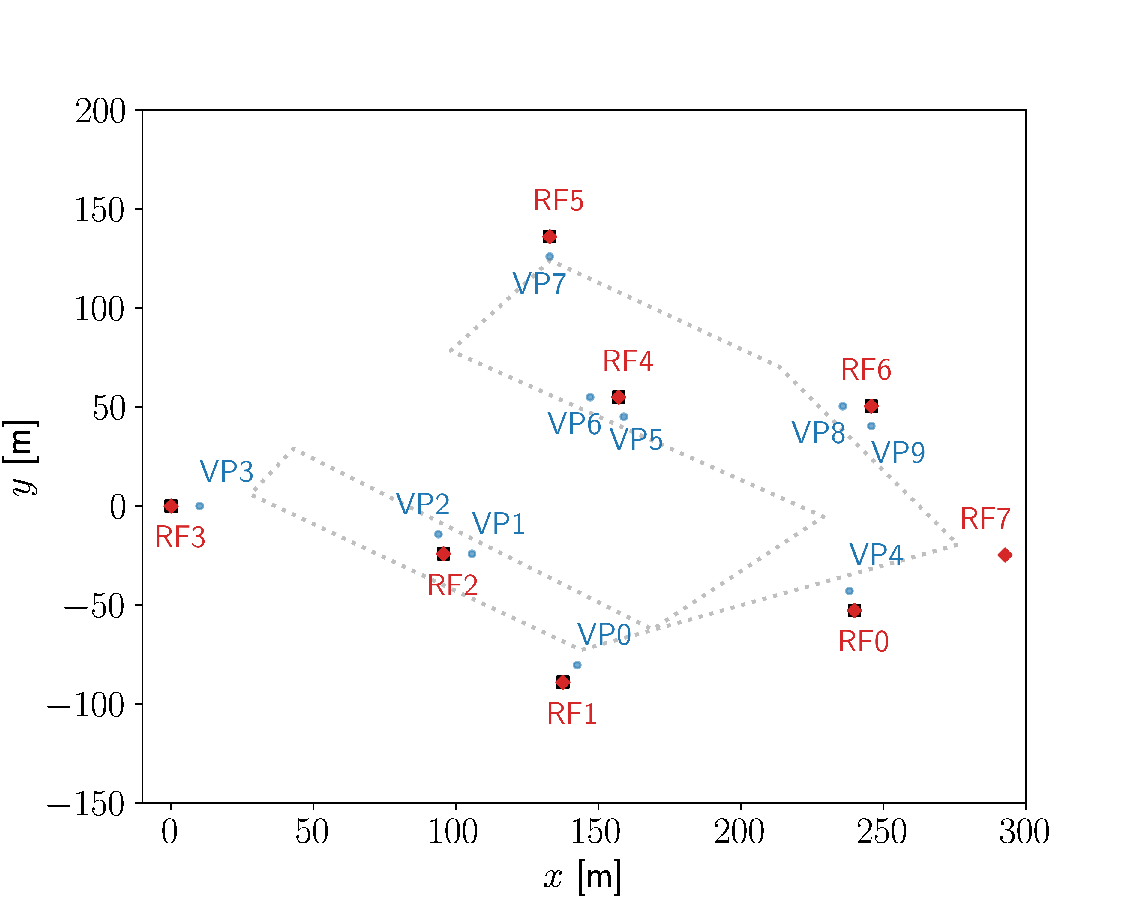}
    \caption{
    Surveillance area with nominal trajectory (gray dotted line). Red boxes denote landmarks, blue dots the viewpoints $\mathcal{VP}$, and blue boxes the corresponding regions $\mathcal{F}(\mathbf{q}_\mathrm{VP}^{(j)})$. Black diamonds mark the \ac{rf} anchors.}
    \label{fig:complete:scenario}
\end{figure}

\begin{figure}[!ht]
    \centering
    \subfloat[\label{fig:complete_239:traj}]{
        \includegraphics[width=\linewidth]{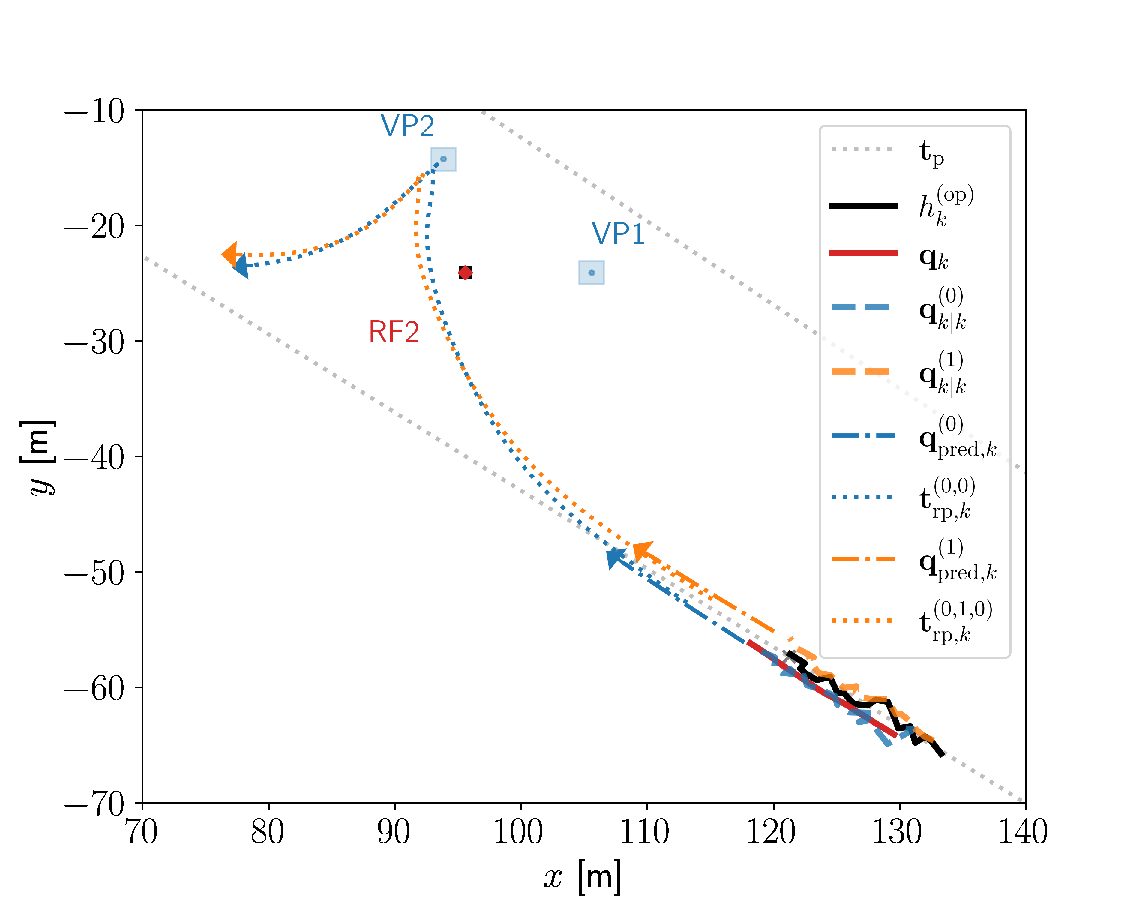}
    }\\
    \subfloat[\label{fig:complete_239:perf}]{
        \includegraphics[width=\linewidth]{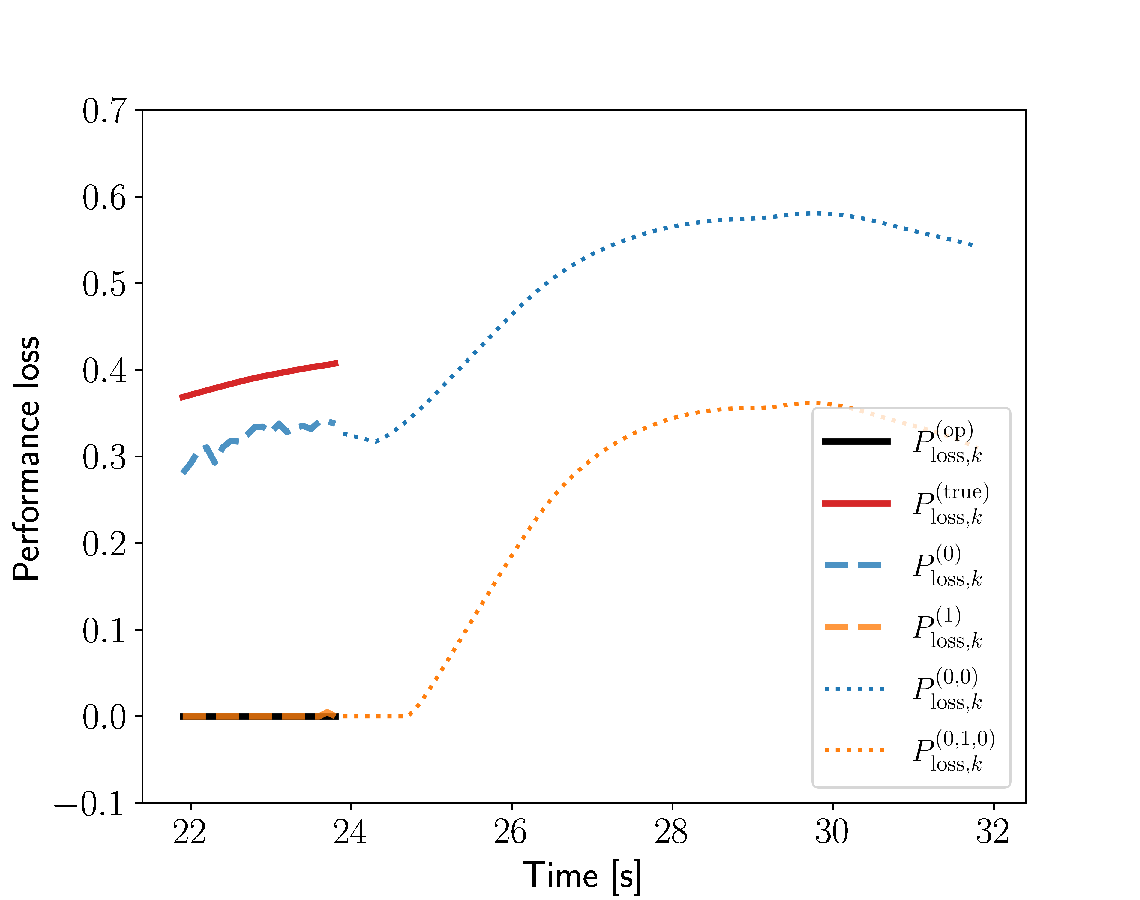}
    }
    \caption{Mitigation phase at time step 196. (a) Re-planned trajectories; (b) performance losses. Supports: $\mathcal{O}^{(0)}=\{\mathrm{GNSS},\mathrm{RF1}\}$, $\mathcal{O}^{(1)}=\{\mathrm{RF0},\mathrm{RF2}\}$.}
    \label{fig:complete_239}
\end{figure}

The second case study demonstrates the full detection and mitigation process. In this scenario is the outlier percentile $\alpha_\chi=95.45\%$, the Poisson binomial percentile $\beta = 99.9\%$ and the similarity percentile $\alpha_\mathrm{F} = 9.95\%$. The scenario in Fig.~\ref{fig:complete:scenario} includes attacks on the \ac{gnss} and \ac{rf} anchors 1, 3, 4, 5, 6, and 7, producing a $3\,\mathrm{m}$ bias in the $x$-direction at time step 20. 
The robot transitions to the diagnosis state immediately and enters the mitigation state at time step 186. At time 196, the re-planner selects VP2 for information gathering, as shown in Fig.~\ref{fig:complete_239}. The corresponding predicted performance loss, computed from Eq.~\eqref{eq:performance-loss-index}, indicates a 50\% degradation limit violation due to intermittent anchor visibility. 

\subsection{Discussion}
\label{sec:sec:discussion}

Assumption \ref{assum:known-natur-outlier-prob} impacts the estimate of the outlier probability, as can be seen in Eq. \eqref{eq:outlier-prob-special}. If the estimate of the naturally occurring outliers $p_m$ is larger than the true value, the \ac{tpr} will decrease, likewise, when it is smaller than the true value, the \ac{fpr} will increase. These probabilities can be determined through empirical datasets \cite{Wymeersch2009}, dependent on the environment in which the robot will operate. If the local modes are not well separated from the global mode, the specialization of Eq. \eqref{eq:outlier-prob-general} to Eq. \eqref{eq:outlier-prob-special} is not valid. However, if the specific modes are known, one can compute the outlier probability given that those specific mode. 

Assumption \ref{assum:E=True} implies that the Bernoulli variables in Eq. \eqref{eq:bernoulli-random-var} are independent, such that one may attribute Eq. \eqref{eq:binom_dist} with the Poisson binomial distribution. The indicator $\delta_k^{(\iota,i)}$ depends on both predicted and measured quantities, $\hat{\mathbf{z}}_{k|k-1}^{(\iota,i)}$ and $\mathbf{z}_k^{(i)}$. If $\hat{\mathbf{z}}_{k|k-1}^{(\iota,i)}$ depends on past measurements, temporal correlation arises between consecutive $\delta_k^{(\iota,i)}$. This occurs when the motion model is inaccurate or linearization errors are large, potentially yielding excessive outlier counts. A practical mitigation is to inflate the posterior covariance, enforcing stronger correction from new measurements and maintaining the nominal assumption of independence over the window $W$. Despite this approximation, the Poisson binomial model provides an effective mechanism for identifying deviations from nominal operation, as illustrated by the low \ac{fpr} in Fig.~\ref{fig:sens:scenario1_resFPR_SPLIT}.

The analysis assumes persistent and colluding malicious sources through Assumption \ref{assum:persistent-attack}. If attacks are uncoordinated but persistent, multiple hypotheses are generated, increasing computational cost due to potential additional re-planning. Introducing hypothesis likelihood weights could prioritize more probable hypotheses and limit re-planning complexity. Adaptive attackers seeking optimal deception must remain within hypothesis gates to avoid rejection, thereby restricting their ability to significantly alter spoofing signals.

\subsubsection{Measurement Separation Strategy}
The sensitivity analysis in Section~\ref{sec:case-studies} varied $\alpha_\chi$, $\beta$, $\alpha_\mathrm{F}$, and the bias magnitude. The window length $W$ influences detection probability and latency. A longer window captures slower anomalies but delays detection, while a shorter one increases responsiveness at the cost of potential false positives. The employed rule requiring hypotheses to exist for at least $W$ steps and to contain at least $W/2$ measurements before transitioning to mitigation ensures stability but could be relaxed for faster reaction, albeit at higher \ac{fpr}.

\section{Conclusion}
\label{sec:conclusion}

This paper presented a robotic system architecture for resilient navigation of a mobile robot equipped with \ac{rf}-based sensors, including \ac{gnss}, \ac{uwb}, and 5G \ac{isac}. The framework integrates fault detection concepts into a multi-hypothesis estimation scheme, enabling identification and isolation of attacked measurement sources without assuming prior knowledge of which sensors may be compromised. A state machine governs transitions between operation, diagnosis, and mitigation, ensuring that the robot maintains situational awareness even under adversarial conditions.

A windowed count detector based on a Poisson binomial distribution was employed to identify deviations from nominal operation, providing low false positive rates and high detection reliability in the presence of biasing attacks. Once an anomaly was confirmed, a mitigation strategy using differential flatness and nonlinear model predictive control re-planned the trajectory to gather additional information, minimizing performance loss while recovering accurate state estimates.

The case studies demonstrated the algorithm’s capacity to distinguish truthful from malicious measurements, maintain navigation capability during spoofing attacks, and autonomously re-plan trajectories to re-establish trust in sensor data. The results indicate that the proposed approach can effectively handle persistent and coordinated attacks on multiple \ac{rf} sources.

Future work include extending the framework to multi-robot systems, integrate a hypotheses weight signifying a trust towards the hypotheses and researching path re-planner frameworks that can unify collision avoidance and the constraint of visiting view points for information gathering.

\appendix
\section{Algebraic Simplification of Outlier Probability}
\label{app:1}
Applying the Leibniz rule on Eq. \eqref{eq:average_likely_prob}, considering the scalar case, differentiating with respect to $r$ and after rearranging, we get the following expression
\begin{equation}
    \bar{P}_{\mathrm{in}}'(r) =\frac{1}{2\pi pr^2}\int_{-\infty}^\infty ye^{-\frac{1}{2}\left(\frac{y}{p}\right)^2}\left(e^{-\left(\frac{y-\gamma_{\alpha_\chi}r}{\sqrt{2}r}\right)^2} - e^{-\left(\frac{y+\gamma_{\alpha_\chi}r}{\sqrt{2}r}\right)^2}\right)\mathrm{d}y
\end{equation}
Completing the square in the exponent, we see that we end with an expression equal to the first moment of a Gaussian distribution. The above can the be simplified and integrated with respect to $r$ which gives
\begin{equation}
    \bar{P}_{\mathrm{in}} =\frac{\sqrt{2\pi}np^2}{\pi}\int_0^{r}\frac{e^{-\frac{n^2}{2}\left(1-\frac{p^2}{r^2+p^2}\right)}}{\sqrt{r^2+p^2}}\frac{1}{r^2+p^2}\mathrm{d}r
\end{equation}
Using three change of variables, first $a=p^2+r^2$ and utilizing $a^2=p^2\sec^2(\theta)$, second $u=\sin(\theta)$ where $\cos(\theta)a=p$ and third $t=\frac{n}{\sqrt{2}}u$ we arrive at 
\begin{equation}
    \bar{P}_{\mathrm{in}} = \mathrm{erf}\left(\frac{nr}{\sqrt{2\left(r^2+p^2\right)}}\right)
\end{equation}

\small
\bibliographystyle{elsarticle-num-names} 
\bibliography{references.bib}

\end{document}

%% file: acronyms.tex
\DeclareAcronym{wsn}{
    short=WSN,
    long=Wireless Sensor Network
}
\DeclareAcronym{iot}{
    short=IOT,
    long=Internet of Things
}
\DeclareAcronym{pdf}{
    short=PDF,
    long=probability density function
}
\DeclareAcronym{cdf}{
    short=CDF,
    long=cumulative distribution function
}
\DeclareAcronym{dof}{
    short=DOF,
    long=degrees of freedom
}
\DeclareAcronym{iid}{
    short=i.i.d.,
    long=independently and identically distributed
}
\DeclareAcronym{rdp}{
    short=RDP,
    long=Ramer-Douglas-Peucker
}

\DeclareAcronym{fisst}{
    short=FISST,
    long=finite set statistics
}
\DeclareAcronym{rfs}{
    short=RFS,
    long=Random Finite Set
}
\DeclareAcronym{mot}{
    short=MOT,
    long=multi object tracking
}
\DeclareAcronym{phd}{
    short=PHD,
    long=Probability Hypothesis Density
}
\DeclareAcronym{lmb}{
    short=LMB,
    long=labeled multi-Bernoulli
}
\DeclareAcronym{fov}{
    short=FOV,
    long=field of view
}
\DeclareAcronym{gm}{
    short=GM,
    long=Gaussian mixture
}
\DeclareAcronym{gc}{
    short=GC,
    long=Gaussian component
}
\DeclareAcronym{bc}{
    short=BC,
    long=Bernoulli component
}
\DeclareAcronym{ci}{
    short=CI,
    long=covariance intersection
}
\DeclareAcronym{emd}{
    short=EMD,
    long=exponential mixture
}
\DeclareAcronym{gci}{
    short=GCI,
    long=generalized covariance intersection
}
\DeclareAcronym{mil}{
    short=MIL,
    long=minimum information loss
}
\DeclareAcronym{ga}{
    short=GA,
    long=Geometric Average
}
\DeclareAcronym{aa}{
    short=AA,
    long=Arithmetic Average
}
\DeclareAcronym{md}{
    short=MD,
    long=Mahalanobis Distance
}
\DeclareAcronym{mle}{
    short=MLE,
    long=Maximum Likelihood Estimation
}
\DeclareAcronym{kf}{
    short=KF,
    long=Kalman filter
}
\DeclareAcronym{ekf}{
    short=EKF,
    long=extended Kalman filter
}
\DeclareAcronym{jpda}{
    short=JPDA,
    long=joint probabilistic data association
}
\DeclareAcronym{mht}{
    short=MHT,
    long=multi-hypothesis tracking
}
\DeclareAcronym{da}{
    short=DA,
    long=data association
}

\DeclareAcronym{kld}{
    short=KLD,
    long=Kullback-Leibler divergence
}

\DeclareAcronym{ospa}{
    short=OSPA,
    long=optimal sub-pattern assignment
}
\DeclareAcronym{gospa}{
    short=GOSPA,
    long=generalized optimal sub-pattern assignment
}
\DeclareAcronym{nees}{
    short=NEES,
    long=normalized estimation error squared
}
\DeclareAcronym{rms}{
    short=RMS,
    long=root mean square
}

\DeclareAcronym{cv}{
    short=CV,
    long=Constant Velocity
}

\DeclareAcronym{aoa}{
    short=AOA,
    long=angle of arrival
}
\DeclareAcronym{aod}{
    short=AOD,
    long=angle of departure
}
\DeclareAcronym{toa}{
    short=TOA,
    long=time of arrival
}
\DeclareAcronym{los}{
    short=LOS,
    long=line of sight
}

\DeclareAcronym{isac}{
    short=ISAC,
    long=integrated sensing and communication
}
\DeclareAcronym{ris}{
    short=RIS,
    long=reflective intelligent surface
}
\DeclareAcronym{bs}{
    short=BS,
    long=base station
}

\DeclareAcronym{nmpc}{
    short=NMPC,
    long=nonlinear model predictive control
}
\DeclareAcronym{qp}{
    short=QP,
    long=quadratic program
}
\DeclareAcronym{fdi}{
    short=FDI,
    long=False Data Injection
}
\DeclareAcronym{dm}{
    short=DM,
    long=Data Modification
}
\DeclareAcronym{glrt}{
    short=GLRT,
    long=Generalized Likelihood Ration Test
}
\DeclareAcronym{fd}{
    short=FD,
    long=Fault Diagnosis
}
\DeclareAcronym{mcs}{
    short=MCS,
    long=Monte Carlo Simulation
}
\DeclareAcronym{rng}{
    short=RNG,
    long=Random Number Generator
}
\DeclareAcronym{fpr}{
    short=FPR,
    long=false positive rate
}
\DeclareAcronym{tpr}{
    short=TPR,
    long=true positive rate
}
\DeclareAcronym{dos}{
    short=DOS,
    long=denial-of-service
}
\DeclareAcronym{cps}{
    short=CPS,
    long=cyber-physical system
}

\DeclareAcronym{gnss}{
    short=GNSS,
    long=global navigation satellite system,
}
\DeclareAcronym{gps}{
    short=GPS,
    long=global positioning system,
}
\DeclareAcronym{rf}{
    short=RF,
    long=radio frequency
}
\DeclareAcronym{uwb}{
    short=UWB,
    long=ultra-wide band
}
\DeclareAcronym{imu}{
    short=IMU,
    long=inertial measurement unit
}

\DeclareAcronym{uav}{
    short=UAV,
    long=unmanned aerial vehicle
}

\DeclareAcronym{mip}{
    short=MIP,
    long=mixed integer programming
}

\DeclareAcronym{ml}{
    short=ML,
    long=machine learning
}

%% file: Figures/GraphicalProof.tex
\begin{tikzpicture}
\begin{axis}[
  no markers, domain=-8:8, samples=100,
  axis lines*=center, xlabel=$x$, ylabel=$y$,
  every axis y label/.style={at=(current axis.above origin),anchor=south},
  every axis x label/.style={at=(current axis.right of origin),anchor=west},
  height=6cm, width=12cm,
  xtick=\empty, ytick=\empty,
  enlargelimits=false, clip=false, axis on top,
  grid = major
  ]
  \addplot[fill=red!20, draw=none, domain=-1.5:0.5]{gauss(0,1)} \closedcycle;
  \addplot [very thick,red] {gauss(0,1)};
  \addplot [very thick,blue] {gauss(0,2)};

  \addplot[blue!80, dashed] coordinates {(-0.5, 0) (-0.5, {gauss(0,2)})};
  \addplot[mark=*] coordinates {(-1, {gauss(0,1)})};
\draw[<-, thick] (axis cs: -0.5, \gaussianY) -- (axis cs: -4.2, \gaussianY+.05)
    node[above] {$\ell_{N\left(0,p\right)}(x)$};
\draw [decorate, decoration={brace, amplitude=10pt, mirror}, thick]
  (axis cs:-1.5, -0.01) -- (axis cs:0.5, -0.01)
  node [midway, yshift=-18pt] {$\mathcal{E}_{r}^{\gamma_\alpha}(x)$};
\end{axis}
gauss(0,1)

\end{tikzpicture}